\documentclass[twocolumn,twocolappendix]{aastex631} 
\usepackage[utf8]{inputenc}
\usepackage{enumerate}
\usepackage{graphicx}
\usepackage{hyperref}
\usepackage{mathtools}
\usepackage{comment}
\usepackage{tabularx}
\usepackage{amsmath}
\usepackage{changepage}
\usepackage{amssymb}
\usepackage{lipsum} 
\usepackage{float}
\usepackage{commath}
\usepackage{enumerate}
\usepackage{longtable}
\usepackage{stackengine}
\usepackage{xcolor}
\usepackage{cleveref}
\usepackage[caption=false]{subfig}
\usepackage{placeins}
\usepackage[ruled,lined]{algorithm2e}

\AfterEndEnvironment{figure}{\noindent\ignorespaces}

\definecolor{darksalmon}{rgb}{0.91, 0.59, 0.48}

\begin{document}
\title{Testing Dark Matter with Generative Models for Extragalactic Stellar Streams}

\author[0000-0001-8042-5794]{Jacob Nibauer}
\altaffiliation{NSF Graduate Research Fellow}
\altaffiliation{Co-first author: these authors contributed equally to the research.}
\affiliation{Department of Astrophysical Sciences, Princeton University, 4 Ivy Ln, Princeton, NJ 08544, USA}
\correspondingauthor{Jacob Nibauer}
\email{jnibauer@princeton.edu}

\author[0000-0003-0256-5446]{Sarah Pearson}
\altaffiliation{Co-first author: these authors contributed equally to the research.}
\affiliation{DARK, Niels Bohr Institute, University of Copenhagen, Jagtvej 155A, 2200 Copenhagen,  Denmark}
\correspondingauthor{Sarah Pearson}
\email{sarah.pearson@nbi.ku.dk}

\shortauthors{Nibauer \& Pearson}

\begin{abstract}
\noindent
Upcoming ground and space-based surveys are poised to illuminate low surface brightness tidal features, providing a new observable connection to dark matter physics. 
From imaging of tidal debris, the morphology of stellar streams can be used to infer the geometry of dark matter halos. In this paper, we develop a generative approach, \texttt{X-Stream}, which translates stream imaging into constraints on the radial density profile of dark matter halos—from the inner region out to the virial radius. Using the GPU-accelerated code \texttt{streamsculptor}, we generate thousands of stream realizations in trial gravitational potentials and apply nested sampling with a custom objective function to explore viable regions of parameter space. We find that multiple stellar streams can be used to constrain the entire radial density profile of a halo, including both its inner and outer density slopes. These constraints provide a test for alternatives to cold dark matter, such as self-interacting dark matter, which predicts cored density profiles. From cosmological simulations, the outer density slope is expected to correlate with merger histories though remains underexplored observationally.
With ongoing and upcoming missions such as Euclid, the Rubin Observatory, ARRAKIHS, and the Nancy Grace Roman Space Telescope, \texttt{X-Stream} will enable detailed mapping of dark matter for thousands of galaxies across a wide range of redshifts and halo masses. 

\vspace{1cm}
\end{abstract}

\section{Introduction}
The nature of the dark matter particle, e.g., its free-streaming velocity, its mass, and whether it self-interacts, affects the overall properties of dark matter halos. Under  
$\Lambda$CDM, the density profiles of dark matter halos is expected to follow a Navarro–Frenk–White profile (NFW; \citealt{nfw1996}), with a steeply rising cusp at the center and a characteristic falloff in the outskirts.
Self-interacting dark matter (SIDM) models predict that the inner density profile of halos is cored with a constant density, while the outer part follows an NFW profile \citep{spergel2000}. 
Warm dark matter models can produce cores in their centers, but some fine-tuning is required \citep[e.g.,][]{strigari2007,maccio2012} to match the observed cores in some galaxies.  
Fuzzy dark matter models predict a central soliton core, resulting in a flat inner density slope, while the outer halo transitions to an NFW-like profile \citep{hui2017}. Therefore, mapping the variations in density slopes can provide insights into the nature of dark matter.

Observations of stellar kinematics and HI in gas rich dwarfs and spiral galaxies show examples of near linearly rising rotation curves, which implies that they have cored density profiles in their centers \citep[e.g.,][]{flores1994,moore1994,gentile2004,oman2015,oh2015,lelli2016,read2017,2025arXiv250722155H}.
The fact that some dwarf spheroidal galaxies host globular clusters at present day can also be used to argue in favor of cored density profiles, as cored profiles would allow for globular clusters to remain close to their currently observed locations for long times instead of in-spiraling due to dynamical friction
 \citep[e.g. in Fornax:][]{goerdt2006,cole2012}. $\Lambda$CDM predicts steeply rising cusps at the center of galaxies, thus these observations pose a challenge to the theory.

\citet{williams2025} recently suggested a theoretical model based on statistical mechanics which shows that dark matter halos can have a diverse range of density profiles, including cored density cores at their centers. The discrepancy between their model and the lack of  cores in dark matter only $\Lambda$CDM $N$-body simulations remains a mystery. 
Including baryons to DM simulations shows that galaxies could have more cored profiles even in $\Lambda$CDM due to bursty star formation and stellar feedback 
\citep[see e.g.,][]{oman2015,straight2025}, but the diversity of radial profiles of dark matter halos remain challenging to reproduce in simulations. 
Inferring cores from rotation curves involves assumptions about gas pressure \citep{pineda2017}, and while   
alternative methods of inferring density slopes without assumptions about dynamical equilibrium exist \citep[e.g.,][]{nguyen2023},  the sample sizes of measured cored density profiles remain low.

The outer halo density profile (i.e., beyond the halo scale radius) depends on the overall amount of accretion experienced by a halo over the past few billion years \citep{diemer2014}. Outer halo slopes are also shaped by filamentary accretion from the cosmic web \citep[e.g.,][]{zhang2013,laigle2025, 2025ApJ...988..190A} and show a correlation with varying environment,
where isolated halos have shallower outer density slopes \citep{avila1999}. 
Although steepening of outer halo profiles can be detected with weak-lensing analyses of massive clusters \citep{diemer2014}, measuring the outer radial profiles of individual halos remains a challenge. 

Stellar streams offer an independent gravitational tracer of dark matter halos. 
When stellar streams form, stars escape from their progenitors due to the underlying mass profile of the host galaxy. The subsequent trajectories of stars around their host galaxy is typically dominated by the dark matter halo properties of the host \cite[]{johnston2001}. Stellar streams thus encode properties of dark matter halos.
Studies of stellar streams in the Milky Way have shown that streams are sensitive to the local acceleration field \citep{bonaca2018,nibauer2022, nibauer2025}, and the shape of the dark matter halo \citep[e.g.,][]{koposov2010,lawmajewski2010,pearson2015,kuepper2015,bovy2016, yavetz2021, yavetz2023,woudenberg2024}. We are now entering an era where it will be possible to extend studies of the Milky Way's dark matter halo to thousands of other galaxies with upcoming low surface brightness observations from {\it Euclid} \citep{racca2016}, The Rubin Observatory \citep{ivezic2019}, The Nancy Grace Roman Space Telescope ({\it Roman}; \citealt{spergel2015}), and ARRAKIHS \citep{guzman2022}. 

One of the first detections of an extragalactic stellar stream was the stream north of M83 shown in \citet{malin1997} and discussed in \citet{bergh1980}. 
We now know of more than 150 extragalactic streams from imaging individual galactic stellar halos \citep[e.g.,][]{delgado2008,mcconnachie2009,denja2016}, and from larger surveys such as DESI \citep{delgado2023}, DES \citep{miro2024}, SAGA \citep{miro2024}, HSC \citep{kadofong2018}, and MADCASH \citep{carlin2016, carlin2019}). Already we are in a regime where the number and diversity of tidal features is applicable to dynamical modeling. 

\citet{Fardal2013} demonstrated that, in the case of the Giant Southern Stream (GSS) in M31—originally detected by \citet{ibata2001}—it is possible to constrain both the mass of M31's dark matter halo and the mass of the GSS progenitor using Bayesian sampling of the parameter space, with each sample evaluated through an $N$-body simulation. However, this approach is computationally expensive and difficult to scale to a larger sample of galaxies, especially those with more limited surface brightness, distance, and kinematic data.

\citet{pearson2022b} developed an extragalactic stream-fitting code that explored a grid of 2D velocity vectors for the progenitor across ten halo mass bins using particle spray simulations \citep{Fardal2015}. They evaluated model accuracy by comparing stream tracks to observed control points and ensured realistic tidal stripping using a Jacobi radius criterion. Results were validated with follow-up $N$-body simulations \citep{kawata2003}. Applying their method to a stream near Centaurus A \citep{denja2016}, they showed that a single progenitor radial velocity can constrain the dark matter halo mass ($M_{200} > 4.7\times 10^{12}~M_{\odot}$). Additionally, radial velocity and distance gradients along the stream offer further mass constraints. However, a more exhaustive parameter search over both progenitor and halo properties is still needed, and was not performed in \citet{pearson2022b} due to computational limitations of their approach.

However, in most cases, we only have access to observations of the projected morphology of extragalactic streams. 
\citet{nibauer2023} showed that the observed morphology of extragalactic streams even without kinematics is sensitive to the shape of dark matter halos. Their technique relies on the curvature distribution of a stream, which can be used to determine a range of plausible halo geometries, particularly the flattening value and direction of the flattening axis. \citet{nibauer2023} applied their technique to the stream enclosing NGC 5907, and ruled out large parts of the allowed parameter space for halo flattening. Their method is not, however, capable of constraining the radial profile of spherical dark matter halos.  

In spherical dark matter halo potentials, where the orbital plane itself does not precess, the precession per orbit between apocenter or pericenter is determined by both the orbital properties (e.g., ratio of radial to azimuthal periods; \citealt{johnston2001,binney2008,hendel2015}), and the slope of the potential \citep{belokurov2014}. 
\citet{walder2025} recently showed that for an orbit launched at apocenter and viewed at a 0 degree inclination, multiple wraps of the orbit can constrain the radial profile of the dark matter halo. 

Building on intuition from previous studies, we develop a generative method, \texttt{X-Stream}, to infer the full radial profile of dark matter halos. The method is built upon the GPU accelerated code \texttt{streamsculptor} \citep{Nibauer2025a}, which leverages the \texttt{JAX} Python framework \citep{Jax2018} to efficiently simulate thousands of streams. From an image alone, the solution space of viable model parameters is highly degenerate due to the lack of kinematic information and projection effects. To sample high-probability islands and capture physical degeneracies, we develop a custom sampling technique to accelerate the inference process and reveal important physical degeneracies. Our method can recover the true underlying radial density profiles in both NFW and cored halos from the on-sky morphology of streams at different radial locations in the potential. 

The paper is organized as follows: In \S\ref{sec:intuition}, we provide an overview of stellar stream evolution in potentials with different radial profiles. In \S\ref{sec:methods}, we introduce our method and lay out the details of the sampling technique. In \S\ref{sec:results}, we show the results of applying our method to distinct streams on different orbits, in \S\ref{sec:discussion} we discuss our findings, and in \S\ref{sec:conclusion} we summarize and conclude.

\section{Stream morphologies as a probe of radial profiles}\label{sec:intuition}
\begin{figure}
\centering\includegraphics[scale=.58]{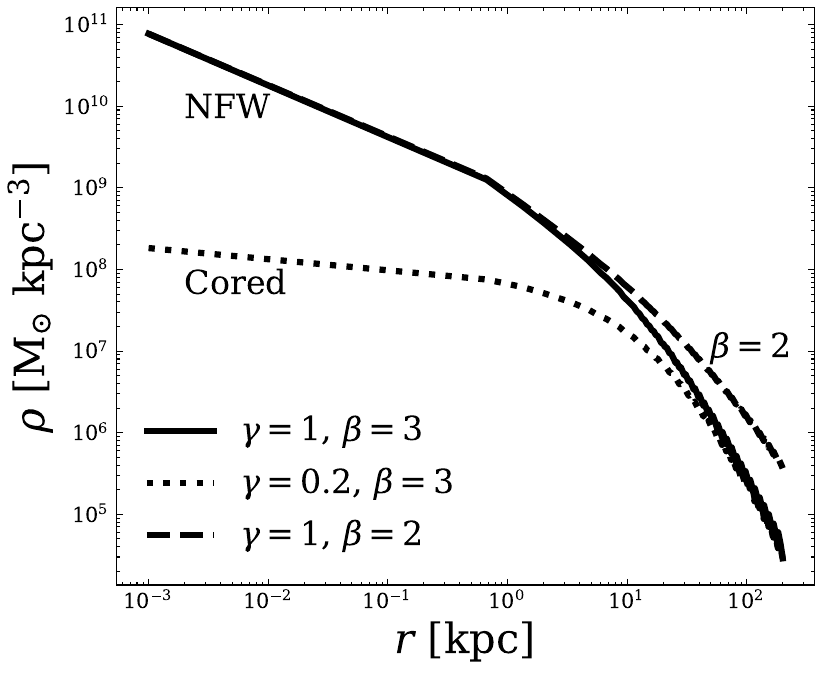}
    \caption{Three different dark matter halo density profiles: NFW (solid line), inner core and outer NFW (dotted line), and inner NFW and outer shallower slope (dashed line). We evolve streams in these three potentials to test their  sensitivity to changes in the radial profiles.
    }
    \label{fig:Core_Cusp_Density}
\end{figure}

The morphology of a stellar stream is characterized by its curvature, width, and length. These properties are determined by a combination of the underlying potential of the host galaxy and the progenitor's orbit. In this work, we investigate dark matter halos with densities of the form:

\begin{equation}\label{eq: gnfw}
\rho(r) = \frac{\rho_0}{(r/r_s)^\gamma (1+r/r_s)^{\beta - \gamma}} \exp\left\{ - \left(\frac{r}{r_{\rm cut}} \right)^2\right\}
\end{equation}
with:
\begin{equation}\label{eq: scalemass}
\rho_0 = \frac{M_{\rm halo}}{4 \pi r_s^3}\frac{1}{{\rm ln}(1 + c_{\rm NFW}) - (c_{\rm NFW}/(1+c_{\rm NFW}))}, 
\end{equation} 
where $M_{\rm halo}$ is the halo scale mass, $r_s$ is the scale radius of the potential, $c_{\rm NFW}$ is the halo concentration, $\gamma$ is the slope of the inner density profile, and $\beta$ is the slope of the density profile beyond the scale radius. The exponential cutoff ensures a finite mass. We use $r_{\rm cut} = 25 \times r_s$, and solve Poisson's equation on a radial grid to infer an interpolated potential from the density. This choice of $r_{\rm cut}$ is not crucial to our modeling, since all streams have apocenters much lower than the minimum $r_{\rm cut}$ used (i.e., for $r_{s,\rm{min}} = 10~\rm{kpc}$, $r_{\rm cut} = 250~\rm{kpc}$). In Fig.~\ref{fig:Core_Cusp_Density} we show examples of three different dark matter halo density profiles: (1) An NFW profile \citep{navarro1997} with $(\gamma$,$\beta)$ = ($1, 3$; solid line), (2) a cored profile, with $(\gamma$,$\beta)$ = ($0.2, 3$; dotted line), and (3) an inner slope following an NFW profile, but a shallower outer slope $(\gamma$,$\beta)$ = ($1, 2$; dashed line).

\begin{deluxetable*}{lcccccccccc}
\tablecaption{Parameters for the inner and outer stream}
\tablecolumns{11}
\tablenum{1}\label{tab:streams}
\tablewidth{0pt}
\tablehead{
  \colhead{Stream} & 
  \colhead{\bf $x$} & 
  \colhead{\bf $y$} & 
  \colhead{\bf $z$} & 
  \colhead{\bf $v$} & 
  \colhead{\bf $\theta$} & 
  \colhead{\bf $\phi$} & 
  \colhead{m$_{\rm prog}$} & 
  \colhead{\bf $t_{\rm age}$} & 
  \colhead{$r_{\rm peri}$} & 
  \colhead{$r_{\rm apo}$} \\
  \colhead{} & 
  \colhead{[kpc]} & 
  \colhead{[kpc]} & 
  \colhead{[kpc]} & 
  \colhead{[km/s]} & 
  \colhead{[deg]} & 
  \colhead{[deg]} & 
  \colhead{log$_{10}$[M$_\odot$]} & 
  \colhead{[Myr]} & 
  \colhead{[kpc]} & 
  \colhead{[kpc]} 
}
\startdata
Inner  & -20 & 30  & 40 & 273.8 & 58   & 280   & 8.0 & 5000 & 20, 32, 14 & 54, 54, 54  \\
Outer  & -40 & 100 & 100& 342.2 & 85.9 & 68.75 & 8.0 & 5000 & 56, 76, 29 & 175, 202, 152 \\
\enddata
\tablenotetext{}{Note that $r_{\rm peri}$ and $r_{\rm apo}$ are listed for the six different streams shown in Fig.~\ref{fig: Inner_Density_TwoStream} evolved in the three different potentials shown in Fig.~\ref{fig:Core_Cusp_Density}. The first values for each stream are for the fiducial NFW halo.}
\end{deluxetable*}

\begin{figure*}
\centering\includegraphics[scale=.65]{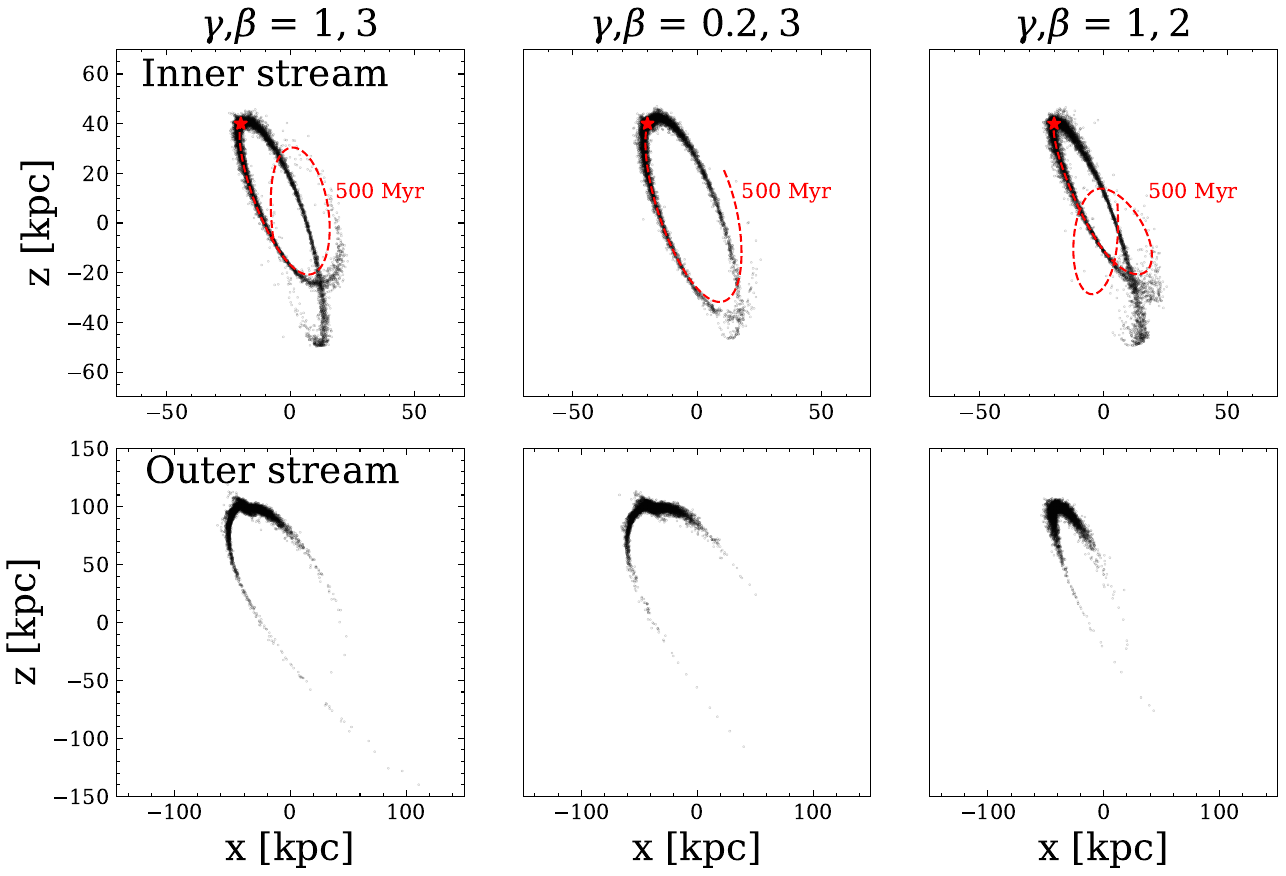}
    \caption{Two different streams evolved for 5 Gyr in the three potentials shown in Fig.~\ref{fig:Core_Cusp_Density}. The  left column displays streams evolved in the NFW profile, the middle column shows streams evolved in a cored profile, and the right column shows a shallower outer halo profile compared to NFW expectations. In the top row the red star marks the progenitor, and dashed red curves denote the progenitor's past orbit over 500~Myr. The streams have differences in their overall morphologies in terms of length, curvature, turning point angles, and radii of their turning points.}
    \label{fig: Inner_Density_TwoStream}
\end{figure*}

To illustrate the sensitivity of streams to variations in halo density profiles, in Fig.~\ref{fig: Inner_Density_TwoStream} we show two different streams which we label the inner stream (top row) and the outer stream (bottom row).
The streams are evolved in the three different potentials shown in 
Fig.~\ref{fig:Core_Cusp_Density} (left: NFW, middle: cored, right: shallower outer profile), with a halo scale mass defined in Eq. \ref{eq: scalemass} of $M_{\rm halo} = 10^{13}$ M$_{\odot}$ (corresponding to $M_{\rm enc}$ at the outer stream's radius of $\approx 4 \times 10^{12}$), $r_{\rm s,halo} = 22$ kpc, $c_{\rm NFW} = 15$, including a disk \citep{Miyamoto1975} with $m_{\rm disk} = 5\times 10^{10}$ M$_\odot$, $a_{\rm disk} = 3$ kpc, $b_{\rm disk} = 0.2$ kpc. Our halo mass is higher than typical spiral galaxies in the local group, but compatible with lower limits from Centaurus A \citep{pearson2022b} \footnote{Note that the density structure of faint dwarf galaxies ($M_* \sim 10^{6}$ M$_{\odot}$) would be of particular interest, as stellar feedback in such galaxies would not modify the central structure \citep{bullock2017}.}. We list the initial conditions and properties of each of the streams in Table \ref{tab:streams}. 
Note that throughout the paper, we fix the disk parameters, but we vary the halo parameters (see Table \ref{tab:freeparams}). 
The red stars in the upper panels show the progenitor location, and the red dashed line shows the past 500 Myr of evolution for each progenitor. Note that the streams were evolved for 5 Gyr. 

Throughout the paper, we use the inner and outer streams as our fiducial stream models to mimic two accreted dwarf galaxies around an external galaxy. The outer stream is short and probes the outer halo region ($r \gtrsim 3 r_s $), beyond the scale radius, and the inner stream is longer and probes the intermediate regions of the potential ($r \lesssim 3 r_s$). For $\gamma=1$ and $\beta = 3$ (NFW), the pericenter of the inner (outer) stream is $20.3~\rm{kpc}$ ($56.6~\rm{kpc}$) and the apocenter is $54~\rm{kpc}$ ($175.3~\rm{kpc}$). The coordinates $x,z$ define the plane of the sky, and $y$ is the line-of-sight direction towards the observer. 
We simulate the streams in the frame of the host galaxy, which has its center at $(x,z)$ = (0,0). We assume that we know the distance to the host galaxy and therefore work in the host galaxy's galactocentric frame. We define the progenitor velocities, which we also list in Table \ref{tab:streams}, as:
\begin{equation}
\begin{split}
\hat{v}_x &= \sin(\theta) \cos(\phi), \\
\hat{v}_y &= \sin(\theta) \sin(\phi), \\
\hat{v}_z &= \cos(\theta), \\
{\bf v} &= v \hat{v}.
\end{split}
\end{equation}

In Fig.~\ref{fig: Inner_Density_TwoStream}, for the inner stream in the top right panel the progenitor completes more radial and azimuthal revolutions for a fixed time of 500 Myr (see red dashed line), i.e. the stream phase mixes more rapidly in the right case.  
\citet{amorisco2015} showed that for steeper inner density slopes (larger $\gamma$), phase-mixing timescales become faster (see their eq. 16-18). This can also be understood by considering the mass enclosed by a stream as a function of $\gamma$, reading off the density profile in Fig.~\ref{fig:Core_Cusp_Density}. For steeper inner densities, the mass interior to the stream is higher and dynamical times become shorter. This allows for more efficient stream growth, as can be seen from the first two columns of Fig.~\ref{fig: Inner_Density_TwoStream}. As $\gamma$ goes from $1\xrightarrow{}0.2$, the streams become shorter and possess different turning points. The effect is most noticeable for the inner stream, which is sensitive to the inner density profile. 

Similar arguments can be made for the outer density slope, $\beta$. In particular, at $r>>r_s$ the density profile goes like $\rho \propto r^{-\beta}$. This is the same asymptotic behavior as for the inner density ($\rho \propto r^{-\gamma}$), though with a different power-law index. Thus, the arguments of \citet{amorisco2015} are applicable to the outer regions of the halo. In terms of density, a larger $\beta$ more strongly truncates the outer density, typically decreasing the enclosed mass. This means the progenitor's orbit will progress more slowly at larger $\beta$. This can be seen in going from the rightmost to leftmost column of Fig.~\ref{fig: Inner_Density_TwoStream}. Namely, for $\beta$ increasing from $2 \xrightarrow{} 3$, the progenitor sweeps out less of its orbit for fixed integration time. The effect on the stream is subtle, though the turning point locations of the inner stream's tidal tails differ, and the curvature of both the inner and outer streams are noticeably different (tighter for small $\beta$). We discuss quantitative expectations for the angle between turning points and the stream's curvature as a function of $\gamma$ and $\beta$ below.

\begin{figure}
\centering\includegraphics[scale=.7]{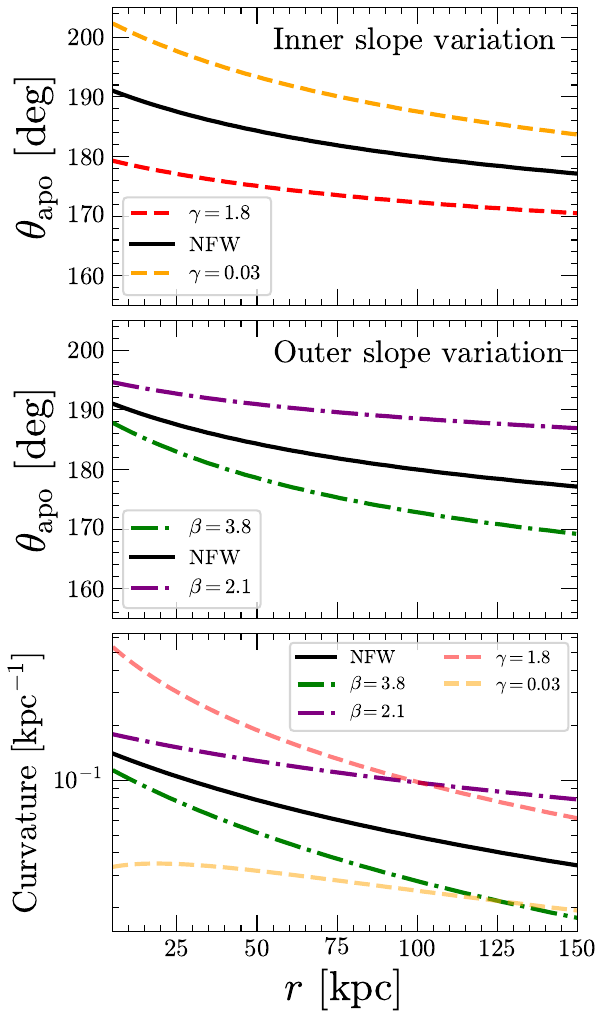}
    \caption{The precession angle between successive apocenters, $\theta_{\rm apo}$, is shown for orbits evolved in potentials with varying inner slopes (top panel) and varying outer slopes (middle panel). In both cases, steeper inner or outer slopes lead to a decrease in $\theta_{\rm apo}$. However, the dominant density slope depends on the galactocentric radius of the orbit. The curvature of an orbit (bottom panel) also depends on the inner and outer density slopes. For increasing inner densities, the orbital curvature increases. For steeper outer densities, the orbital curvature decreases.
    }
    \label{fig: intuition}
\end{figure}

The top two panels of Fig.~\ref{fig: intuition} show the angle between successive apocenters (i.e., the angle between orbital turning points), $\theta_{\rm apo}$, for the streams evolved under the mass model in Eq. \ref{eq: gnfw} with varying $\gamma$ and $\beta$ (see labels) as a function of galactocentric radius. We calculate $\theta_{\rm apo}$ using the epicycle approximation (Appendix~\ref{sec:theory}). In Fig.~\ref{fig: intuition},  steeper inner and outer density slopes (larger $\gamma$ and larger $\beta$) lead to smaller angles between successive apocenters. Note, however, that at small galactocentric radii $\theta_{\rm apo}$ is more sensitive to changes in $\gamma$, whereas at larger galactocentric radii $\theta_{\rm apo}$ is more sensitive to changes in $\beta$ due to the relations between the radial and azimuthal periods (see Appendix \ref{sec:theory}). We can understand the trends between $(\gamma,\beta)$ and $\theta_{\rm apo}$ as follows. As $\gamma$ increases, the inner regions of the potential become deeper and the radial and azimuthal periods both decrease, with their ratio approaching unity. Thus, the angle between apocenters diminishes. For $\beta$, the mass enclosed approximately goes like $M_{\rm enc} \propto r^{3-\beta}$ below the cutoff radius, $r <r_{\rm cut}$, though beyond the scale radius, $r_s$. At large $r$, increasing to $\beta > 3$ the mass distribution becomes more centrally packed-in and orbits become approximately Keplerian, for which $\theta_{\rm apo} \xrightarrow[]{} 0$. Thus, we expect that increasing $\beta$ causes turning point angles to decrease. Indeed, this is what we find in Fig.~\ref{fig: Inner_Density_TwoStream}, and show analytically in Fig.~\ref{fig: intuition}. We note that the angles between turning points are degenerate with projection effects. The generative method we develop in \S\ref{sec:methods} will account for viewing angle degeneracies.

The angles between successive apocenters are easier to visualize in the orbital plane of the sky. In Appendix \ref{sec:theory}, Fig.~\ref{fig:orbitalplane}, we show the three inner streams rotated to the orbital plane in $x,z$ and visualize their two most recent apocenters. As expected from Fig.~\ref{fig: intuition}, the left inner stream has the smallest $\theta_{\rm apo}$, and the right inner stream has the largest $\theta_{\rm apo}$. This is consistent with the findings in Fig. 9 of \citet{belokurov2014} and Fig. 1 of \citet{walder2025}, although note their different choice of potential forms. 

Another observable property of extragalactic streams is the stream curvature \citep{nibauer2023}. The curvature, $\kappa = \Vert \boldsymbol{\kappa} \Vert$, of a stream is defined as:
\begin{equation}\label{eq: curv}
    \boldsymbol{\kappa} = \frac{\Vert \boldsymbol{v} \times \nabla \Phi \Vert}{\Vert \boldsymbol{v}^3\Vert} \hat{N},
\end{equation}
where $\boldsymbol{v}$ is the local velocity of the stream, and the unit vector $\hat{N} \propto d\boldsymbol{\hat{v}}/dt$ is normal to the local mean orbit, along the perpendicular component of the acceleration vector. Intuitively, at fixed velocity a larger force generates more strongly curved orbits and streams. In \citet{nibauer2023}, the observed curvature direction (i.e., $\hat{N}$) was used to infer the geometry of the potential. By using a generative model, we can simultaneously produce a velocity distribution of stars belonging to the stream, and attempt to self-consistently determine the range of potential models that are consistent with the observed curvature, both in its magnitude and direction (whereas in \citealt{nibauer2023}, the direction was used). We demonstrate the sensitivity of orbital curvature to the radial density profile in the bottom panel of Fig.~\ref{fig: intuition}. To generate this figure, we assume a fixed velocity of $220~\rm{km/s}$ perpendicular to the acceleration. When increasing the inner density slope (orange to red dashed lines), the curvature tends to increase. This is because for a more cuspy profile the mass enclosed increases, generating larger forces. When increasing the outer density slope (purple to green lines), the curvature tends to decrease. This is because for larger $\beta$, the outer density becomes more truncated, decreasing the forces interior to the orbit. Note that we have presented a simplified picture here, and that our real sampler will deal with degeneracies induced by projection effects which are intertwined with the observed curvature. However, Fig.~\ref{fig: intuition} illustrates that the geometry of a stream, including the angular spacing of loops and the curvature distribution, contains information about the halo it resides in.

We have demonstrated that two different extragalactic streams at two different locations in their host halo are sensitive to different parts of the potential \citep[as expected from Milky Way studies, e.g.,][]{bonaca2018, nibauer2022}, and that streams, which cross or are within the scale radius, are particularly sensitive to the inner slope of the potential. 
Note that while deviations from spherical symmetry can also introduce orbital plane precession, in this paper we focus on spherical dark matter halos as the simplest mass model that is also interesting for dark matter physics \citep{spergel2000,bullock2017}. 

\section{Method}\label{sec:methods}
Based on our intuition from the previous section,
we now develop a generative method, \texttt{X-Stream}, to constrain dark matter halos from images of stellar streams.

\subsection{Stream Generation and Model Parameters}
To generate stellar streams in different potential models, we use the particle-spray method \citep{Fardal2015} implemented in the GPU accelerated code \texttt{streamsculptor} \citep{Nibauer2025a}, built within the \texttt{JAX} python framework \citep{Jax2018}. The GPU support of our simulator allows us to efficiently generate stream models in batches of $500$, where each stream in the batch is simulated within a different underlying potential. Each model stream consists of 1000 particles, which is sufficient for the low resolution data representation that we will discuss in \S\ref{sec: mock_obs}. Despite the lack of dynamical friction and star-by-star force calculations, the particle spray approach successfully characterizes the tracks of dwarf galaxy streams when compared to $N-$body simulations \citep{Fardal2015,pearson2022b}.

We consider the two streams presented in \S\ref{sec:intuition}, which trace different locations of the same potential.  We aim to investigate whether the full radial profile of the host halo (i.e., the inner and outer density slopes) can be recovered from the morphology of the two streams.
In Table~\ref{tab:freeparams} we list all the free and fixed parameters adopted in our sampling.   

To generate streams, we fix the on-sky location of the progenitors. Several extragalactic streams have tentative progenitor locations (see e.g., \citealt{pearson2022b,2024A&A...691A.196M}). However, future work can add free parameters for the progenitor's on-sky location (e.g., along the ridgeline of the stream's track). We find that incorrectly choosing the angular location of the progenitor biases progenitor parameters, but does not change our conclusions on inferring the full radial density profile (inner and outer) of the halo.
We fit for the line-of-slight location of the progenitor in its host galaxy, $y_{\rm prog}$, along with its present-day velocity direction, sampled on the velocity unit sphere 
(see \S\ref{sec:intuition} and Table \ref{tab:freeparams}). When sampling from our prior, we test whether the progenitor's orbit is bound. We reject parameters that produce unbound orbits.

Priors on the progenitor properties (e.g., $M_{\rm prog}$, $t_{\rm age}$) are identical, except for the inner stream's line-of-sight progenitor location, $y_{\rm prog}$, which has a tighter prior compared to the outer stream (see Table~\ref{tab:freeparams}). This is because a wider prior in $y_{\rm prog}$ for the inner stream led to many unbound orbits in our sampling scheme, so we reduced the prior range to allow for more efficient sampling. For simplicity, we fix the disk parameters because the streams considered have pericenters of at least a factor of 7 times the scale-length of the disk. However, future work can treat (e.g.) the disk mass as a free parameter. These streams are not unique to our methodology, but their radial locations provide a useful test case to determine what information can be gleaned about the full radial profile of dark matter halos from stream imaging. 

To test various dark matter halo potentials, free parameters for the generalized NFW profile include the halo mass, $M_{\rm halo}$, the scale-radius, $r_s$, the inner density slope, $\gamma$, and the outer density slope, $\beta$. Priors on all halo parameters are uniform. For the slope parameters, we allow for symmetric deviations around a spherical NFW profile, for which $\gamma = 1$ and $\beta = 3$. Note that we fix the halo concentration parameter to $c_{\rm NFW} = 15$ since it is degenerate with the scale radius \citep{Koposov2023}.

\begin{deluxetable}{lccc}
\tablecaption{Priors on model parameters}
\tablecolumns{4}
\tablenum{2}\label{tab:freeparams}
\tablewidth{0pt}
\tablehead{\colhead{\bf } & 
\colhead{\bf inner }  &{\bf outer  }  &{\bf unit  }    
}
\startdata
\hline
Free parameters & &&\\
\hline
$y_{\rm prog}$ &   [-150, 150]&  [-200, 200]& [kpc]   \\
$v$  & [0.2, 0.45] &  [0.2, 0.45]& [kpc/Myr]\\
$\theta$  & [0, $\pi$] & [0, $\pi$]&   \\
$\phi$&[0, 2$\pi$]& [0,2 $\pi$]&\\
log$_{10}M_{\rm prog}$&[6.5, 8.5]&[6.5, 8.5]&[M$_{\odot}$]\\
log$_{10}M_{\rm halo}$&[11.8, 14]&[11.8, 14]& [M$_{\odot}$]\\
$r_s$   &  [10, 30]  & [10, 30]&[kpc]  \\
$\gamma$& [0, 2]&[0, 2]\\
$\beta$ &[2, 4]&[2, 4]\\
$t_{\rm age}$ & [3, 7]& [3, 7]&[Gyr]\\
\hline
Fixed parameters& &&\\
\hline
$x_{\rm prog}$ & -20 & -40& [kpc]  \\
$z_{\rm prog}$ & 40 &100& [kpc]   \\
$m_{\rm disk}$&5e10&5e10&[M$_{\odot}$]\\
a$_{\rm disk}$ & 3 &3&[kpc] \\
b$_{\rm disk}$ &0.2&0.2&[kpc] \\
c$_{\rm NFW}$ & 15 & 15 
\enddata
\tablenotetext{}{}
\end{deluxetable}

\begin{figure*}
\centering\includegraphics[scale=.7]{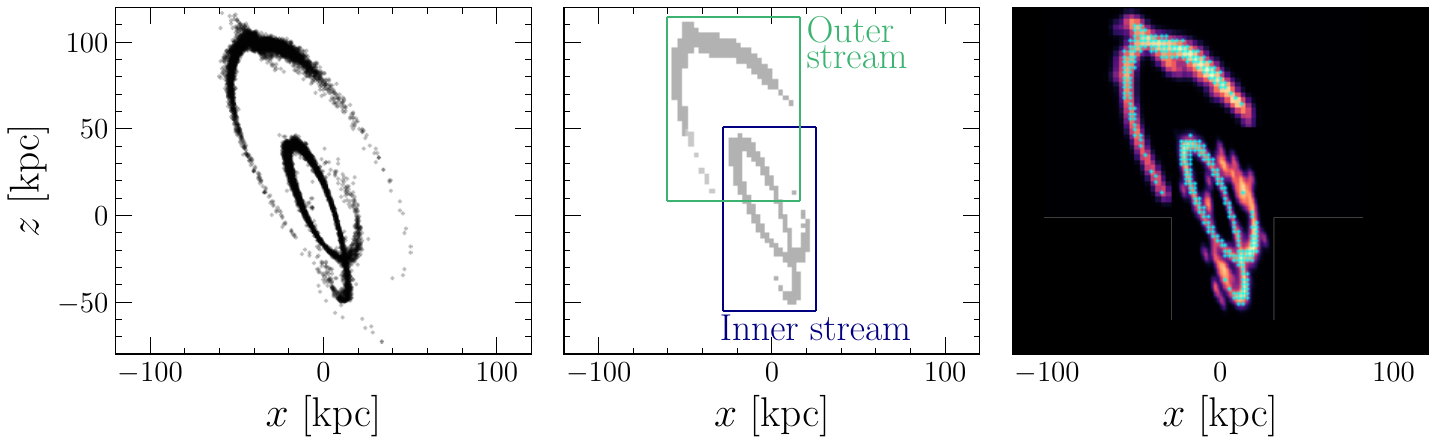}
    \caption{Left: particles of the inner and outer streams evolved in a NFW halo with a disk. 
    Middle: histogram version of our two streams with a threshold applied to represent mock observations of the densest regions of the streams. Right: Control points (cyan) from the gray histograms, over plotted on the corresponding kernel density estimation (KDE) of the mock streams used in our likelihood evaluations. 
    }
    \label{fig: Histogram_Stream}
\end{figure*}

\subsection{Mock Observations}\label{sec: mock_obs}
We create mock observations of the inner and outer streams which we use as the input for our method.  In Fig.~\ref{fig: Histogram_Stream} (left) we show the particle positions of the inner and outer streams (same the left panels of Fig.~\ref{fig: Inner_Density_TwoStream}). 
For current and future stream imaging data (e.g. from Euclid, Roman, or LSST), we have access to the on-sky morphology of stellar streams. However, the faintest parts of the streams will be difficult to detect \citep[see e.g.,][]{bullock2005,miro2024b}. Since we are working with particle data from simulations, we construct histogram representations of the two streams to emulate mock observations (Fig.~\ref{fig: Histogram_Stream}, middle panel). We then place representative control points in each bin that contains a few stars. For the inner stream, we use 30 bins in $x \in [-50,50]~\rm{kpc}$ and $z\in[-60,50]~\rm{kpc}$. The outer stream spans roughly twice the area, so we double the bin size, using $x\in[100,80]$ and $z \in [20,180]$. We apply a threshold mask by setting the count of bins to zero if they contain fewer than $0.2\%$ of the total number particles in each stream's tidal tail. All model streams consist of 1000 particles. However, for the mock data, we increase the resolution: the inner stream contains 2000 particles, and the outer stream contains 8000 particles. The higher particle count is used solely to automate the selection of control points. With too few particles, the control points become sparse and unrepresentative of the actual stream structure, and their distribution becomes highly sensitive to bin size. This limitation does not affect our sampling procedure, as binning is used only to generate mock observations--not for comparing models to data. 

From these histograms, the grid of control points spanning each stream's width and length is shown in the right panel of Fig.~\ref{fig: Histogram_Stream} (cyan points). Specifically, we place $N$ control points in the bin centers along the mock-observed streams, labeled $\{(x_i, z_i)\}$. We assume the same prior probability for each control point, thereby neglecting surface brightness information. Surface brightness profiles could be included in future work. The probability of any given control point is then $1/N$, where $N$ is the total number of control points. 

When we run \texttt{X-Stream}, at every choice of model parameter (see priors in Table \ref{tab:freeparams}), we generate a mock-stream. A bounding box is drawn around the mock observed stream (Fig.~\ref{fig: Histogram_Stream}, middle panel), and we remove any particles from trial models that are outside of the bounding box. This ensures that we do not penalize models that are longer than the observed stream. Note, however, that  model streams shorter than the mock observations are penalized. 

We fit each model with a 2D kernel density estimate (KDE) in a two stage process, in order to smooth over any density features in the mock-stream. 
First, we fit the KDE assuming equal weighting for all model particles. Second, we evaluate the density of the KDE at the location of every model point. We then fit a new KDE, where each particle is weighted by the inverse density at its location. The KDE has a bandwidth of $0.1~\rm{kpc}$. We choose this value because it is within the width of our model streams, and we use stream width in our fits. However, going below this bandwidth can lead to excess Poisson noise in our representation of model streams, making parameter exploration less efficient. Example KDEs for model streams generated at the true input parameters are shown in the right panel of Fig.~\ref{fig: Histogram_Stream}. The result is a smooth ridge-like density representation of each stream that is contained within the bounding box. Note that the density of the extended tails are similar to the regions near the progenitor, in order to avoid penalizing models with differing surface density profiles. To apply the method to observational data, one would represent the observed stream with a grid of control points spanning the stream's width and length.

\subsection{Cost function}
We label the final density estimate of the simulated stream with $p_{\rm model}\left(x_i,z_i\right)$. When evaluating $p_{\rm model}$ over the control points, we find that the  function is very sensitive to the parameters of the potential, because even small changes in progenitor or potential parameters can lead to a misalignment between data and model. This poses a significant challenge for standard sampling techniques, as the underlying likelihood surface is extremely narrow and peaked.  Note however, this indicates that stream morphology is very sensitive to the underlying dark matter halo. To overcome the sampling challenge, we use a technique inspired by likelihood tempering, where the likelihood function is modified (e.g., raised to a power) to reduce extreme sensitivity to the input parameters. This aids in more efficient sampling, though at the cost of a direct interpretation of the posterior distribution. In what follows, we will introduce the cost function to be sampled, and provide a recipe for drawing statistical contours from the output of our sampling procedure.

Our cost function is the Kullback–Leibler (KL) divergence \citep{kullback1951} between the distribution of control points ($p_{\rm data}$) and the model distribution ($p_{\rm model}$). Statistically, one can interpret the KL divergence as an average log-likelihood ratio:
\begin{equation}\label{eq: kl_model}
    {\rm{KL}}\left(\theta\right) \equiv \left\langle \log\left(\frac{p_{\rm data}}{p_{\rm model}}\right) \right\rangle,
\end{equation}
where $\theta$ are model parameters, $p_{\rm data}$ is the prior probability of control points ($1/N$), and the average is over the control points ($x_i,z_i$). We use nested sampling to generate samples from Eq.~\ref{eq: kl_model}. Details of the nested sampling procedure will be discussed in \S\ref{sec: sampling}. Once we generate a distribution of parameters, $\theta$, we must then draw $68\%$ and $95\%$ high confidence regions using the procedure defined in \S\ref{sec: contours}.

\subsection{Sampling}\label{sec: sampling}
From Eq.~\ref{eq: kl_model}, the optimal $\theta$ will minimize the KL divergence. We therefore seek to maximize the function $-\rm{KL}(\theta)$. We accomplish this using nested sampling, implemented in the \texttt{nautilus} python package \citep{nautilus}. Unlike standard nested sampling techniques which involve a rejection step, \texttt{nautilus} employs importance sampling, allowing every proposal sample to contribute to evaluating the posterior distribution. Additionally, the proposal distribution is adaptively learned using multilayer perceptron neural networks, which significantly accelerates the inference process.

While these features of the \texttt{nautilus} code enhance the efficiency of the sampling procedure, the most substantial speedup comes from the vectorized nature of our simulator. Specifically, we use the \texttt{vmap} (vectorized map) functionality in the \texttt{JAX} framework to simultaneously simulate 500 different parameter sets, and thus stream realizations. This parallelization of likelihood evaluations enables much faster exploration of the parameter space, reducing the runtime of our method from over 10 hours to, at most, a few hours. Further speedups may be achieved in future iterations, as discussed in \S\ref{sec:limitations}.

\subsubsection{Drawing Contours}\label{sec: contours}
Having generated samples from $-\rm{KL}(\theta)$, we must now decide which samples belong to regions of $68\%$ and $95\%$ confidence. We will make a number of simplifying assumptions in converting our samples of Eq.~\ref{eq: kl_model} to posterior estimates, and validate the quality of the posteriors in \S\ref{sec:results}. Note that we take a similar statistical approach to \citet{2015ApJ...801...98S}, who used the KL divergence as a test-statistic and then determined regions of 68 and 95\% confidence as a post-processing step. 

First, we identify the best fit model that minimizes Eq.~\ref{eq: kl_model}. The optimal parameters are $\theta_{\rm best}$. Next, we compare all other models against the best fit, through the summary statistic
\begin{equation}\label{eq: lambda_ts}
\begin{split}
    \Lambda &\equiv N\left[ \rm{KL}\left(\theta_{\rm best}\right) - \rm{KL}\left(\theta_{\rm model}\right) \right] \\
    &= \sum_{i=1}^N \log\left( \frac{p_{\rm model}\left(x_i,z_i\right)}{p_{\rm best}\left(x_i,z_i\right)} \right),
\end{split}
\end{equation}
where $p_{\rm best}$ is the KDE of the best fit model. To estimate $X\sigma$ contours, we will approximate $p_{\rm model}$ and $p_{\rm best}$ as Gaussian distributions, and determine the appropriate levels to draw contours based on the ratio of Gaussian density functions with different standard deviations. 
Suppose $p_{\rm best}\left(x_i, z_i\right)$ is approximately Gaussian with zero mean and unit variance in the variable $r_i^2 = x_i^2 + z_i^2$ (e.g., $\mathcal{N}(r_i | 0, 1)$). Because $p_{\rm best}$ represents the best fit, all other models must be wider by a standard deviation factor $\sigma \geq 1$ (e.g., $\mathcal{N}(r_i | 0, \sigma)$). This implies that for two Gaussians,
\begin{equation}
    \Lambda\left(\sigma\right) = -\sum_{i=1}^N \left[ \log(\sigma) + \frac{r_i^2}{2}\left( \sigma^{-2} - 1\right) \right].
\end{equation}
Using the assumption of a standard normal, $\mathbb{E}(r_i^2) = 1$, we find 
\begin{equation}
    \bar{\Lambda}(\sigma) \equiv \mathbb{E}\left(\Lambda(\sigma)\right) = -N\left[\log(\sigma) + \frac{1}{2\sigma^2} - \frac{1}{2} \right].
\end{equation}
Crucially, $\sigma$ represents the deviation from the best fit model, with $\sigma = 1$ representing zero deviation. Therefore, it is convenient to define $\bar{\Lambda}$ in terms of $\Delta \sigma = \sigma - 1$. That is,
\begin{equation}\label{eq: Lambda_bar}
    \bar{\Lambda}(\Delta \sigma) = -N\left[ \log\left(\Delta \sigma + 1 \right) + \frac{1}{2\left(\Delta \sigma + 1 \right)^2} - \frac{1}{2} \right].
\end{equation}
From Eq.~\ref{eq: Lambda_bar} we can find the value of $\Delta\sigma$ corresponding to the $\Lambda$ value calculated for each model (Eq.~\ref{eq: lambda_ts}). That is, we solve for the inverse $\Delta \sigma = F^{-1}\left({\Lambda}\right)$, where $F^{-1}$ is the inverse of Eq.~\ref{eq: Lambda_bar}. We fit for the inverse using spline interpolation. Because $\Delta\sigma$ represents a deviation from the best fit, it is effectively a $z-$score. We therefore can compute the probability of a $\Delta\sigma$ value using the standard normal, $p(\Delta \sigma) \propto \exp\left[-\frac{1}{2} (\Delta\sigma)^2\right]$. This allows us to determine a weight for each sample, which is then used to enclose regions of $68$ and $95\%$ confidence.

In Appendix~\ref{app: test_statistic}, we validate that Eq.~\ref{eq: Lambda_bar} predicts measured values of $\Lambda$ (Eq.~\ref{eq: lambda_ts}) under the conditions of normality. 
In \S\ref{sec:discussion}, we will illustrate that even for more complicated distributions, Eq.~\ref{eq: Lambda_bar} provides a means to enclose regions of parameter space containing samples that match the input data.

We note that because we sample the negative KL-divergence, the density of samples can be different from the true posterior (i.e., the average in Eq.~\ref{eq: kl_model} can suppress sharp peaks in $p_{\rm model}$). To correct for this, posterior constraints are derived by constructing bins in parameter space and taking the average weight of the samples in each bin. Contours are then drawn using the average weight in each bin. This approach ensures that contours are drawn according to the weights rather than the sampling density of the tempered likelihood.

\begin{figure*}
\centering\includegraphics[scale=.32]{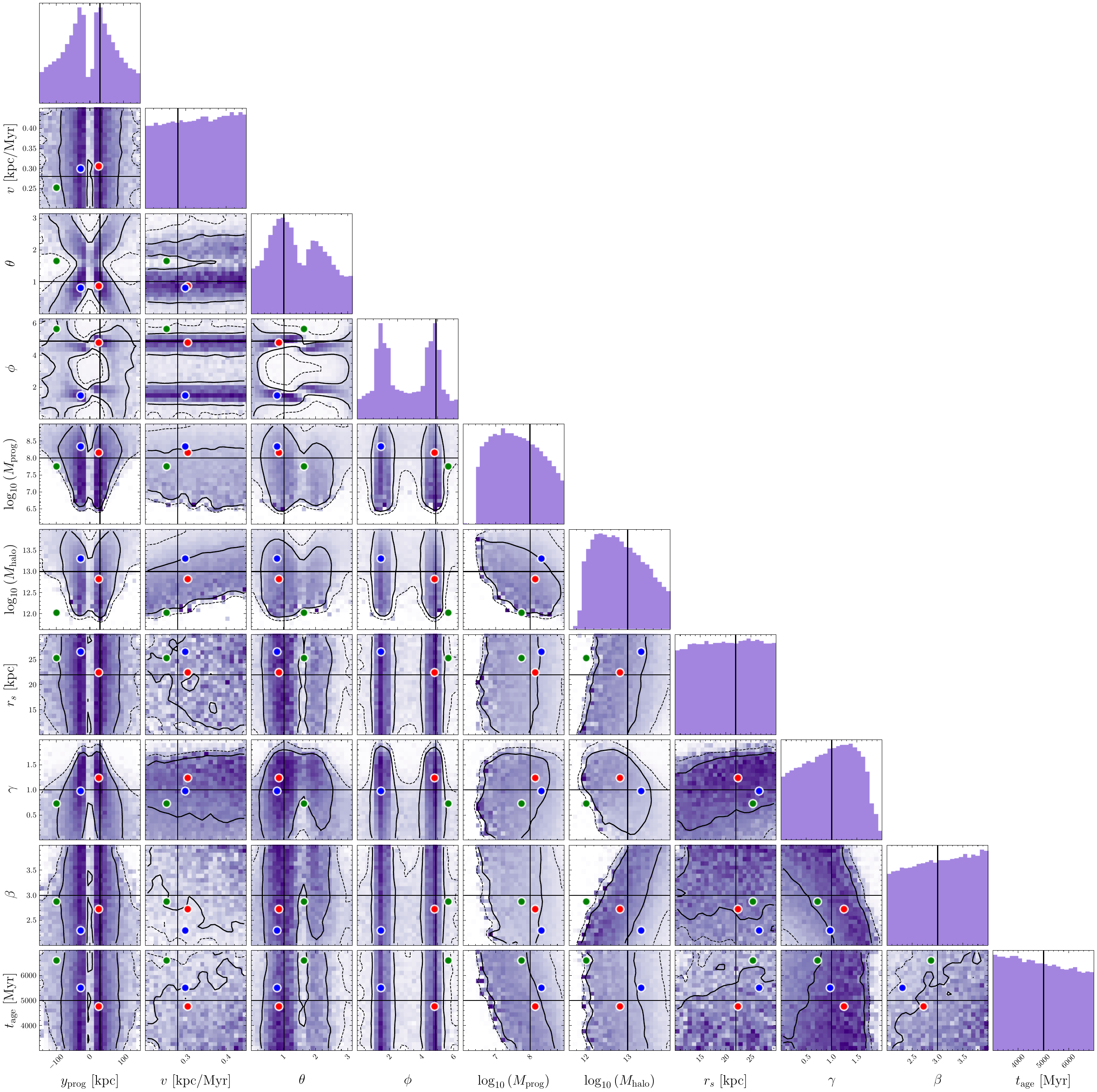}
    \caption{Constraints on the 10 free parameters for the inner stream. The black lines show the true parameters. The red and blue points represent two randomly selected good fits within the 68\% credible region, while the green point corresponds to a poor fit. We visualize the streams corresponding to these points in Fig.~\ref{fig:1sigma3sigma}. The sampler recovers all true parameters within the $68\%$ region, and places strong limits on the inner slope of the density profile, $\gamma$. There are minimal 1D marginal constraints on $v$, $r_s$, $t_{age}$, though these quantities are degenerate with $M_{\rm halo}$, and $t_{\rm age}$ is degenerate with $M_{\rm prog}$. There is information on the line-of-sight distance, $y_{prog}$, and velocity directions, $\theta, \phi$. }
    \label{fig:Corner}
\end{figure*} 

\begin{figure*}
\centering\includegraphics[scale=.65]{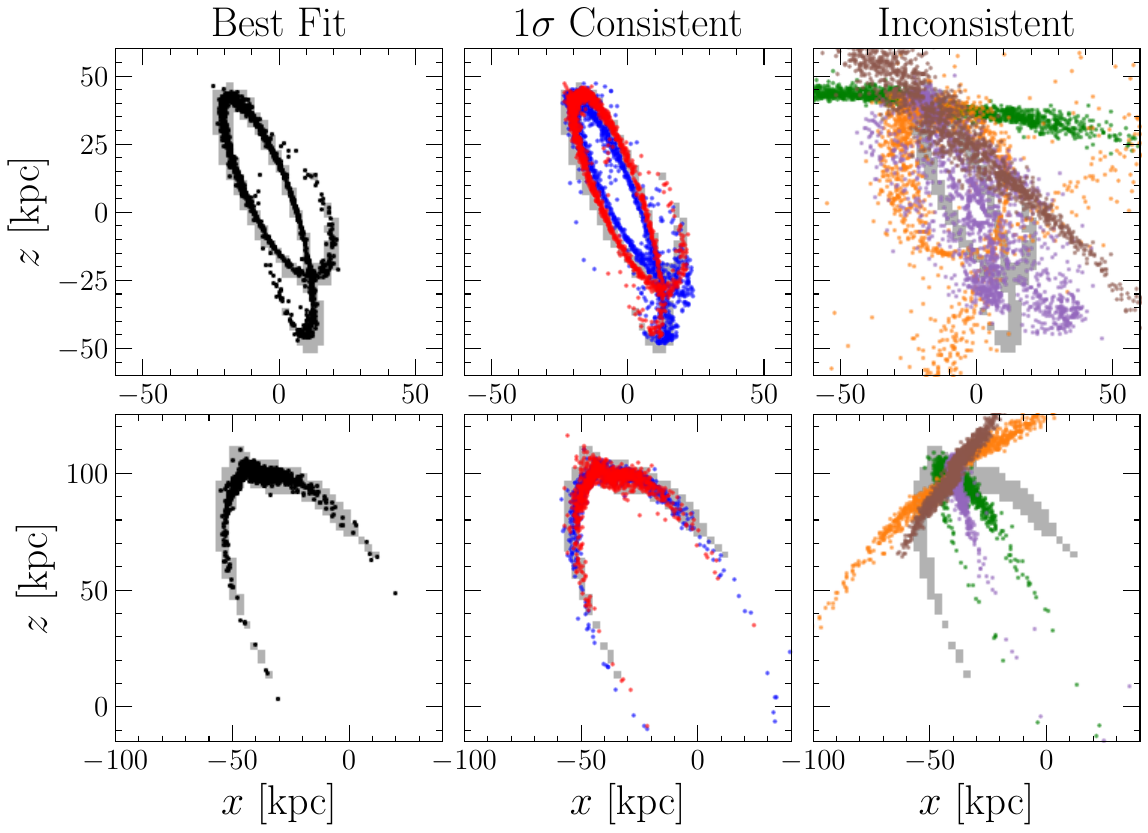}
    \caption{Illustration of the best fit streams (left column), streams within the 68\% confidence region (middle column), and streams deemed to be poor fits (right column). The gray histograms show the input data for the inner stream (top row) and outer stream (bottom row; same as in Fig.~\ref{fig: Histogram_Stream}). Red points represents a stream which resides in front of the host galaxy (closer to the observer) and the blue stream is  behind the host galaxy. The parameters of each stream are indicated by the color-coded points in Fig.~\ref{fig:Corner} (inner) and Fig.~\ref{fig:Corner_outer} (outer).
}
    \label{fig:1sigma3sigma}
\end{figure*}

\section{Results}\label{sec:results}
\subsection{Parameter constraints}
We apply our method to the two streams introduced in \S\ref{sec:intuition}. Our sampler has 10 free parameters and 5 fixed parameters. See Table \ref{tab:freeparams} for the prior ranges for both the inner and outer stream.

In Fig.~\ref{fig:Corner} we present parameter constraints using our method for the inner stream. In Appendix \ref{sec:outer}, Fig.~\ref{fig:Corner_outer}, we present the same constraints for the outer stream. The purple contours show the 68\% and 95\% confidence regions. The true parameters are over plotted as black lines. The colored points represent solutions that fall within the $68\%$ region (blue and red), and outside the $95\%$ region (green). 
 All 10 free parameters for the progenitors and the dark matter halos are recovered within the $68\%$ region. Degeneracies will be discussed in \S\ref{sec:degeneracy}, though we note that for the inner stream, we find a preference for $\gamma = 1$ (the true value), with $\gamma > 1.7$ disfavored at 95\% confidence. Limits on $\gamma$ are placed because the inner stream's apocenter is roughly equal to the halo's scale radius. The inner stream does not rule out any range of $\beta$ (outer slope), though there is a slight preference for values below 2.3 at the 68\% level. For the outer stream, there is no strong preference on the inner density slope, with a 68\% lower limit of $\gamma > 0.3$. For the outer stream, the inference of $\beta$ is slightly better with $\beta>2.4$ at 68\% confidence. The outer stream is less sensitive to $\gamma$ and more sensitive to $\beta$ since its pericenter is around 2.5$\times$ larger than the halo's scale radius.

We also find that the method naturally produces limits on the halo mass ($M_{\rm halo}$). These constraints arise because the mock streams have specific widths, which enters into the stream model through the tidal radius: $r_t\propto \left(\frac{M_{\rm prog}}{M_{\rm enc}}\right)^{1/3}$. Thus selecting a range for the progenitor mass while keeping the stream width fixed will in turn limit the halo mass. For an otherwise fixed progenitor, increasing the halo mass will typically decrease the width of streams. This sets an upper bound on $M_{\rm halo}$ \citep{Erkal2016}, since too massive of a halo will produce a stream narrower than the observations. We also reject unbound progenitors, which further limits the maximum mass of our halo model. Additional constraints come from the observed length of the stream. Streams shorter than the observed length are disfavored. This places a lower limit on the halo mass: if the mass is too small, the resulting stream will be typically shorter at a fixed distance compared to the data. Note, however, that there is also a degeneracy with the dynamical age, $t_{age}$ of the stream (see \S\ref{sec:degeneracy}) and $M_{\rm prog}$.
The stream morphology is not highly sensitive to the scale radius, $r_s$, or to the stream age as indicated by the flat posterior distributions. 
Note that while mass bound can be obtained for the simple particle spray method we have used here, real tidal stripping is more complicated and mass constraints should be viewed as potentially highly model dependent. We defer an exploration of the robustness of mass constraints to future work.

To demonstrate that the sampler successfully identifies orbital solutions and halo profiles consistent with the mock observations, we present in Fig.~\ref{fig:1sigma3sigma} the maximum likelihood forward models for the inner stream (top left) and the outer stream (bottom left). Model streams are compared to the input data (gray histograms), and show excellent agreement.
The middle panels of Fig.~\ref{fig:1sigma3sigma} display stream models corresponding to the blue and red points (typical points from the reweighted samples) and green points (inconsistent) from Figures \ref{fig:Corner} and \ref{fig:Corner_outer}. We also plot other examples of inconsistent streams that are not supported by our sampling in Fig.~\ref{fig:1sigma3sigma} (orange and brown models). The blue model corresponds to a scenario in which the progenitor is located behind the host galaxy at the present day (i.e., negative $y_{prog}$), while the red model has the progenitor in front of the host galaxy (i.e., positive $y_{prog}$). Both of these models closely match the input data. In contrast, the green, brown, and orange models shown in the rightmost column provide a poor fit to the input data as expected. Note we have illustrated one degeneracy here (line-of-sight distance), additional degeneracies will be discussed in \S\ref{sec:degeneracy}.

\begin{figure}
\centering\includegraphics[scale=.5]{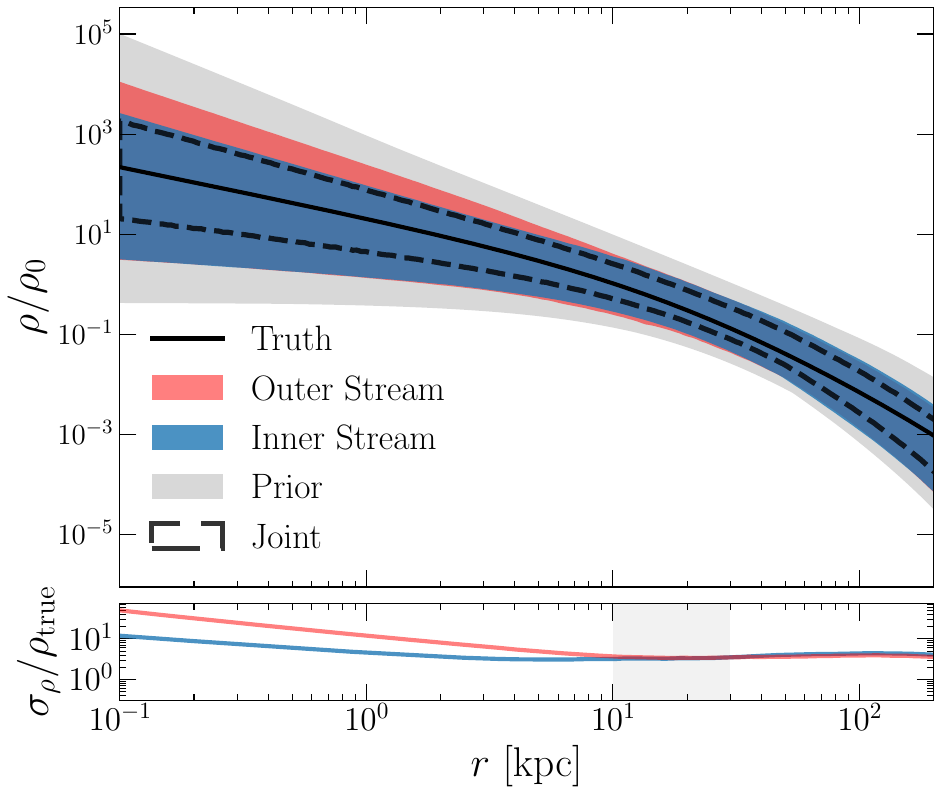}
    \caption{Top:  Constraint on the radial density profile from the inner (blue), outer (red), and combined (black dashed curves) stream samples. The true profile is NFW (black line), and the gray region shows the allowed density profiles from our priors. Both streams show comparable constraints on the outer slope, as their orbits are typically a factor of $2\times$ the halo scale radius. However the inner stream places tighter constraints on the inner slope, since its orbit reaches the scale radius of the halo. Bottom: fractional uncertainty on the radial profile of the halo as a function of distance. The inner stream's errorbar is a factor of $\approx 5 \times$ tighter at small radii, while the outer stream offers marginally tighter constraints on the outer density profile compared to the inner stream. The gray shaded region shows the prior range on the scale radius, $r_s$.
  }
    \label{fig: DensityProfile_Constraint}
\end{figure}

In Fig.~\ref{fig: DensityProfile_Constraint} we summarize constraints on the full radial profile of the halo using both the inner and outer streams. The true density profile is plotted in black, normalized to the scale density $\rho_0$. We isolate our constraints on the shape of the radial profile by plotting the range of models consistent with our fits, normalized by the true scale density. We overplot constraints from the inner stream in blue and the outer stream in red (both at 68\% confidence).
The gray region shows the range of priors for our sampled parameter space over ($r_s, \gamma, \beta$). We have validated that draws from the priors are distinct from the inferred parameter distribution, since the former is flat (i.e., uniform) and the latter is peaked (i.e., prefers a narrower range of values). The dashed region illustrates the constraint from a joint fit. To generate the joint fit, we assume the distribution of normalized densities at each distance slice is approximately Gaussian, and we take the product of Gaussian densities. This treatment of combing information is justified because we have flat priors over all model parameters, so we are effectively combining information at the likelihood level assuming the streams are statistically independent. 

In the lower panel of Fig.~\ref{fig: DensityProfile_Constraint} we plot the fractional uncertainty in the inferred density relative to the true value. 
The inner stream provides a constraint that is approximately five times tighter in the inner regions of the halo compared to the outer stream. Both streams yield similar constraints on the outer slope, though the outer stream performs slightly better. The minimum uncertainty for the inner stream occurs around $15~\rm{kpc}$, while for the outer stream, it occurs around $26~\rm{kpc}$. The shaded gray region in the bottom panel denotes the prior range for the scale radius, $r_s$. We highlight this region because the minimum uncertainty on the density profile is expected to fall in this range, because the scale-radius marks the transition region between the inner density power-law and the outer density power-law. Still, even within the prior range, the radial location of the outer stream's best constraint is at a distance approximately twice of the inner stream's best constraint. This supports the intuition that each stream constrains the region of the potential local to its orbit (e.g., \citealt{bonaca2018, nibauer2022}), with closer-in streams providing more information on the inner regions of the halo.

\begin{figure*}
\centering\includegraphics[scale=.85]{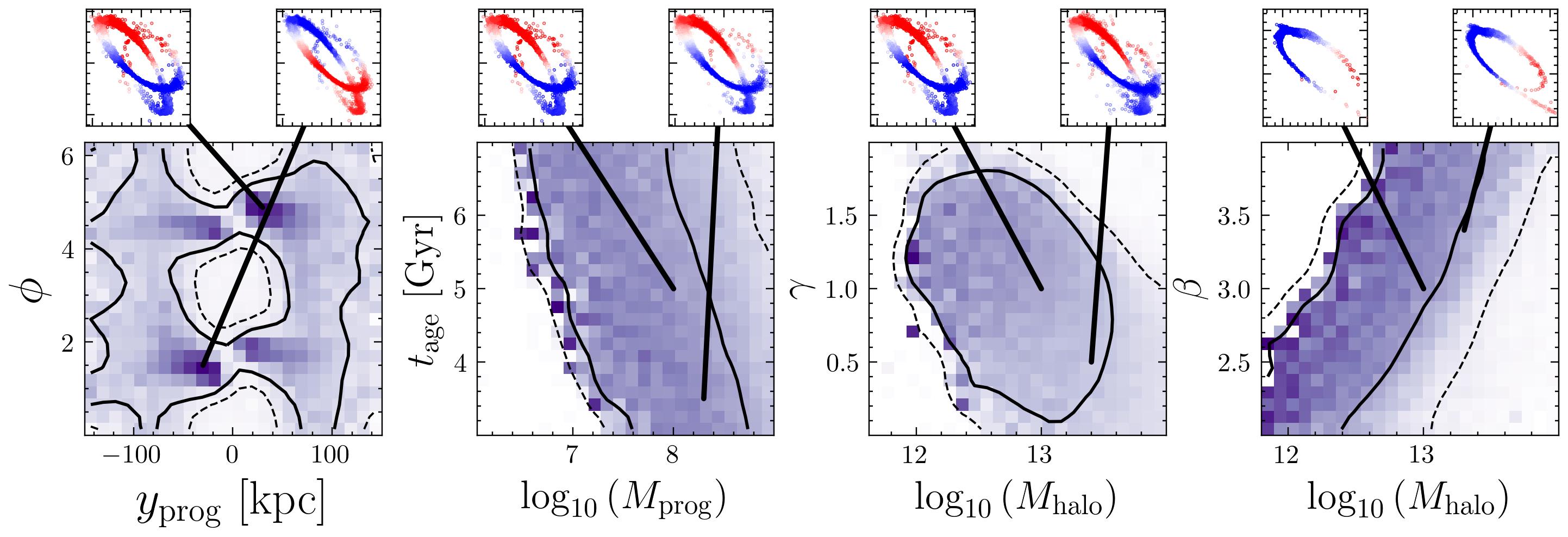}
    \caption{Degeneracies recovered using our method on the inner stream. 
    First panel: The y-axis shows the $\phi$ velocity, which is the azimuthal velocity angle from the $x-$axis, that controls how much of the velocity is along the line-of-sight. The x-axis shows the line-of-sight location, $y_{prog}$, of the progenitor (i.e. negative value mean in front of the host galaxy while positive values mean farther away). 
    In the top panel we show two examples of streams colored by radial velocity (blue: moving towards the observer, red: moving away from the observer) generated from distinct peaks in the likelihood surface. The radial velocity changes sign for the leading and trailing stream in the two cases, but the streams otherwise appear identical.  
    Second panel: streams with larger progenitor masses need lower integration times to reproduce the length and width of the input data. Third panel:  a steeper inner radial profile, $\gamma$, allows for more rapid phase mixing, which means the stream requires a lower halo masses to reproduce the morphology of the input data. Fourth panel: a stepper outer radial profile, $\beta$, requires a higher halo mass to match the input data, since larger $\beta$ truncates the outer regions of the density profile. For two randomly selected peaks within the high likelihood regions for panels 2-4, we do not find large differences in the radial velocity distribution of the streams. 
     }
    \label{fig:degeneracy}
\end{figure*} 

\subsection{Degeneracies}\label{sec:degeneracy}
In Fig.~\ref{fig:degeneracy}, we illustrate some of the degeneracies for the inner stream. 
The panels show examples of two streams sampled within the $68\%$ region, colored by line-of-sight velocity. In the left panel we show the angle $\phi$, which controls how much of the progenitor's velocity is along the line-of-sight, versus $y_{\rm{prog}}$, the progenitor's line-of-sight location. The bimodality in $\phi$ shows the leading or trailing tail can be swapped without impacting the fit, as expected, since there is not a clear ``S"-shape for either stream in our mock-observations. The modes are separated by roughly $\pi$ radians, which is also the case for the outer stream (see Fig.~\ref{fig:Corner_outer}). 
We plot two streams from each of the peaks in the likelihood surface (see Fig.~\ref{fig:degeneracy} top left panels), and recover the \citet{pearson2022b} result (see their Fig. 10), which shows a flipped radial velocity sign for a stream which has its progenitor in front of or behind the host in line-of-sight distance. Red indicates positive radial velocity (moving away from the observer), and blue indicates negative radial velocity (moving towards the observer).

In the second panel of Fig.~\ref{fig:degeneracy} we highlight the degeneracy between the progenitor mass, $M_{\rm prog}$ and the dynamical age, $t_{age}$. For a more massive progenitor, the particle-spray method produces a larger scatter in velocity as stars escape the Lagrange points (see \S\ref{sec:methods}), which leads to more rapid phase mixing thereby producing longer streams. This means that the same length stream can be achieved by increasing the progenitor mass, while decreasing the dynamical age. The two samples illustrated from the high probability regions are extremely similar in radial velocity and morphology.

In the third panel of Fig.~\ref{fig:degeneracy}, we show the inner slope, $\gamma$, halo mass degeneracy, and in the fourth panel we show the outer slope, $\beta$, halo mass degeneracy. 
For higher $\gamma$, we see a degeneracy that correlates with lower halo mass. This confirms our intuition, as larger $\gamma$, i.e. a steeper inner density slope, leads to more rapid phase mixing and thus longer streams (see \S\ref{sec:intuition}). To produce a stream with the same length, we in turn need to lower the halo mass. 
In the fourth panel, we see that a steeper outer slope (higher $\beta$) leads to a degeneracy correlating with higher halo mass. This also confirms our intuition that  higher $\beta$ truncates the intermediate-outer regions of the density, so a higher overall mass amplitude is needed to match the input data (see Fig.~\ref{fig:Core_Cusp_Density}).
The two streams randomly sampled within the 68\% region of the likelihood surface do not show differences in morphology or radial velocity (see top panels).

Note that there are also known degeneracies between progenitor mass, the progenitor orbit, the host dark matter halo mass, and the dynamical age of the streams. \citet{johnston2001} presented a semi-analytic framework to predict the morphology and dynamical age of stellar streams in a logarithmic dark matter halo, which was validated against $N$-body simulations. One of their results was that massive halos will lead to more rapid phase mixing and therefore longer streams. Thus, to produce the same stream morphology in a more massive halo, the integration time needs to be shorter to compensate \citep[see][Eq. 13]{johnston2001}.
Despite some differences in the setup of our stream models (e.g., our streams evolve in a NFW halo), we confirm the intuition built in \citet{johnston2001}: if we did not allow for a long enough integration time in our priors, the method preferred higher halo masses in order to reproduce long enough streams.

\subsection{Cored density profiles}\label{sec:core}
We have showed that  our method can recover the inner and outer slope of a NFW profile. Here we test whether our approach can also recover the true potential parameters if the stream is evolving in a dark matter halo with a cored density profile. Cored profiles are particularly interesting for dark matter physics, since they are expected outcomes in self-interacting dark matter \citep{spergel2000,bullock2017}. As the outer stream was not as sensitive to the inner slope ($\gamma$) of the potential (see Fig.~\ref{fig: Inner_Density_TwoStream}), we attempt to recover $\gamma$ with the inner stream. To run our test, we simulate the inner stream in a cored potential with $\gamma = 0.2$ and $\beta = 3$.

We rerun our sampler for the inner stream over the same 10 free parameters as described in \S\ref{sec:results}, where we only fix the on-sky projected position of the progenitor ($x,z$), as well as the stellar disk mass, scale length, and scale height.  In Fig.~\ref{fig: inner_cored}, left, we present the best fit stream (black) and the control points used for our sampler (red). In the right panel, we show the recovered inner and outer slope parameters (black lines are the input values), which are both recovered within the $68\%$ region. There is a preference for a non NFW profile at the $2\sigma$ level. We visualize consistent streams from our sampling in Appendix \ref{sec:app_core}.

Thus, for streams with orbits crossing interior to the scale radius of the host's dark matter halo, we can use stream imaging to test whether individual galaxies prefer cored or cusped geometries, without the need for kinematics. This result is consistent with \citet{walder2025}, who showed that the logarithmic slope of a power-law potential can be inferred with orbit fitting. Here, we do not assume a single stellar orbit and arrive at the same conclusion with a generative model for the stream's track and width. Combined with recent findings that SIDM can leave distinctive imprints on the evolution of stellar streams  \citep{hainje2025}, streams can provide a powerful test of SIDM.

\section{Discussion}\label{sec:discussion}
In this section, we discuss
how our method compares to other approaches for modeling extragalactic tidal streams (\S\ref{sec:curvature}), and discuss the limitations of our method (\S\ref{sec:limitations}).

\subsection{Comparison to Prior Work}\label{sec:curvature}
\citet{johnston2001} showed that the length, width, and disruption rate of a stream is intertwined with progenitor properties and the characteristics of the host galaxy. Their approach successfully models the properties of streams in a logarithmic halo potential. We have reproduced the degeneracies discussed in \citet{johnston2001}, (e.g., stream length, width) and extend their initial work to an automated method of characterizing the radial profile of dark matter halos and the progenitor's properties.

\citet{pearson2022b}, developed a method to model extragalactic stellar streams using the location of their on-sky tracks. In their work, the 2D velocity vector of the progenitor is varied over several bins in halo mass. They identify new degeneracies including those between the line-of-sight distance with the host and the velocity of the progenitor. Our method is a natural extension to their work, as we perform a much larger search over a higher dimensional parameter space, including the full 3D velocity of the progenitor's orbit, its mass, and halo properties. We recover the same  degeneracies first highlighted in \citet{pearson2022b}, and account for new degeneracies between inner and outer density slopes and halo parameters. Our method also accounts for stream width, while in \citet{pearson2022b} the stream track is modeled. 

\begin{figure*}
\centering\includegraphics[scale=.5]{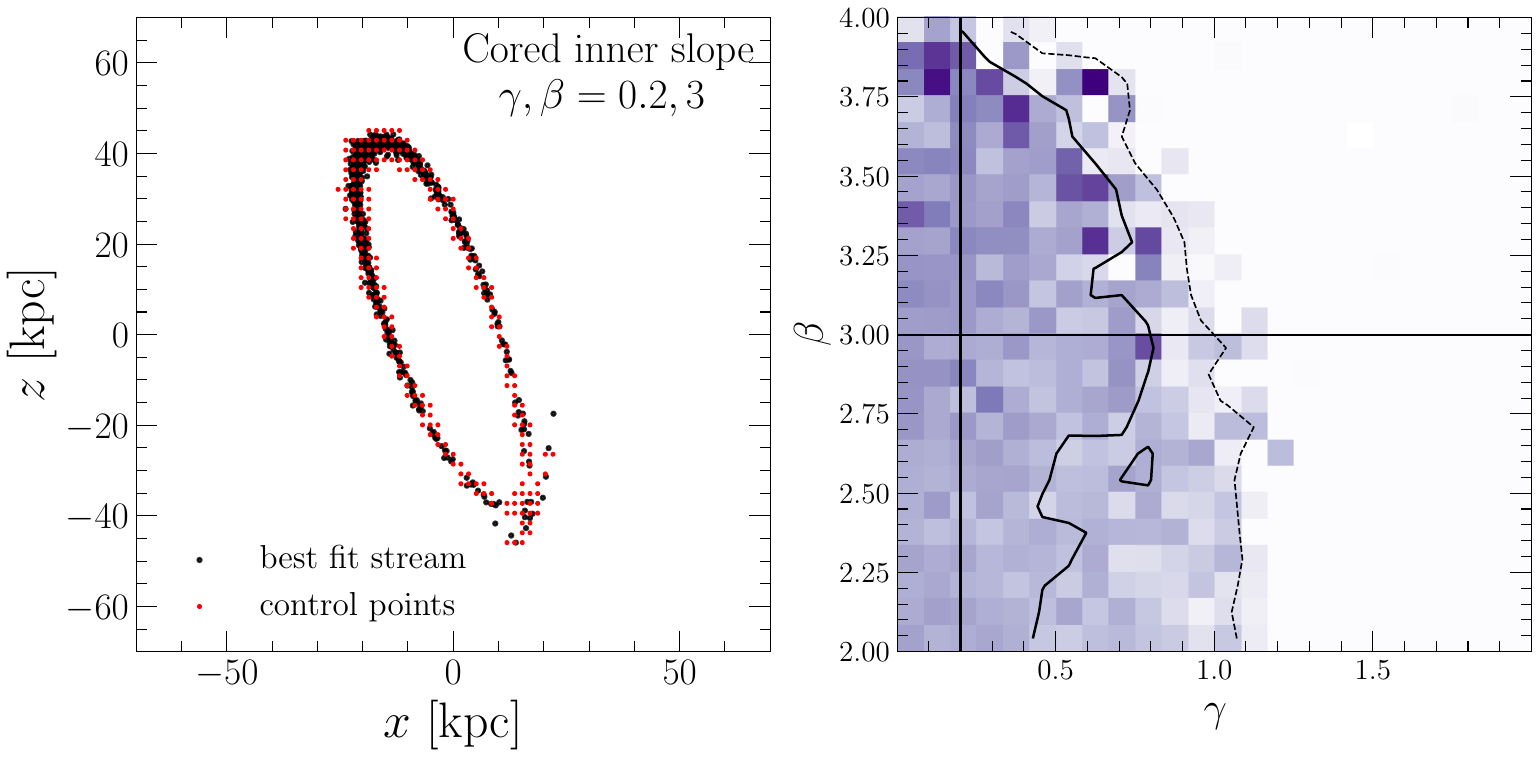}
    \caption{Left: black points show the best fit inner stream evolved in a cored halo with $\gamma, \beta= 0.2,3$.  The red points show the input data used to represent the inner stream. Right:  constraints on  $\gamma$ and  $\beta$. The true values of both parameters are shown as black lines. There is a preference for a cored profile. }
    \label{fig: inner_cored}
\end{figure*} 
\citet{nibauer2023} presented an analytic method that uses the curvature direction of extragalactic stellar streams to constrain the geometry of dark matter halos (e.g., flattening, disk-halo misalignment angles). By comparing the stream's local curvature to projected acceleration fields, the method identifies compatible halo flattenings without modeling full orbits--enabling rapid inference even without a known progenitor. However, their approach cannot constrain spherical halo radial profiles, as it only uses the curvature direction ($\hat{N}$ in Eq.~\ref{eq: curv}), not the magnitude. In contrast, our approach models the full phase-space distribution of stream particles, allowing us to constrain the full $\boldsymbol{\kappa}$ curvature vector from Eq.~\ref{eq: curv}. This makes our generative method sensitive to the radial profile of the potential. However, this is at a cost. First, we must make assumptions about the stream's formation and its time evolution in the halo over several Gyr. Second, while we have taken measures to increase the speed of our method, it is orders of magnitudes slower than the pure curvature approach and not yet applicable to tens of thousands of galaxies. While slower, our method complements the curvature approach. Ultimately, we aim to combine both: using curvature to pre-select systems where a spherical halo is plausible, or to fix flattening parameters inferred from the curvature method for more detailed modeling of radial profiles using \texttt{X-Stream}. We defer this hybrid strategy to future work.

\citet{walder2025} developed a method to fit orbits to extragalactic stellar streams in a power-law potential of the form $\Phi \propto r^{-\gamma}$. They fit orbits to mock-streams launched from apocenter as a function of different inclination angles. We have reproduced their findings that it is possible to infer the inner slope of the potential (or density) from the turning points of a stream's track. We have improved upon their method in two main regards. First, we do not fit orbits to the stream's track, but use a full generative stream model to compare to the observations. Because streams can be significantly misaligned with their progenitor's orbit, orbit fitting can lead to bias \citep{2013MNRAS.433.1813S}. Additionally, our approach includes stream width and length information as constraints on the potential. We find that in combination, these can be used to set mass bounds on the halo. Second, our model has more flexibility to model the underlying stream and dark matter halo. In particular, we model the progenitor's mass and age, while also allowing for variations in an inner and outer halo density slope. We find that multiple streams can be used to stitch together the full radial profile. The disadvantage of our approach is that it is likely slower than the method from \citet{walder2025}, because we compute $1000$ orbits per stream realization. It is worth considering whether a hybrid technique that uses less orbit integrations while also retaining the flexibility we highlight above could lead to further improvements in speed (e.g., the streakline approach \citealt{kuepper2012}).

\subsection{Limitations}\label{sec:limitations}
Throughout this paper, we have made some assumptions to optimize our sampling scheme. Here we discuss some of the limitations of the method.

    {\bf Uniform stripping times}: 
    To simulate mock extragalactic dwarf streams, we have used the particle-spray technique \citep{Fardal2015} implemented in \texttt{streamsculptor} \citep{Nibauer2025a}. This allows us to rapidly generate mock stellar streams on different orbits in various potentials, where stars are 
    released primarily from the Lagrange points uniformly in time. 
   
    For dwarf galaxies, however, much of the stripping occurs near pericenter \citep[see e.g.][]{bonaca2025}. 
    In \citet{pearson2022b}, they included a Jacobi radius stripping criterion for their dwarf progenitor, which ensured that there was no tidal stripping if the Jacobi radius was much larger than the extent of the progenitor. This in turn led to preferred stripping near pericenter, where the Jacobi radius is smaller. Adding a Jacobi radius criterion to our method would only make it more robust and able to reject more orbits. We can relax the assumption of uniform stripping similar to that of \citet{pearson2022b} in future work.
    
    {\bf Progenitor location}: 
    Throughout this paper we fix the present day on-sky position of the progenitor ($x,z$). In observations we will often not have access to the progenitor's sky position, which introduces some uncertainty to our method \citep[e.g.,][]{delgado2023,miro2024}. In future work, we can overcome this by marginalizing the progenitor's location over the stream's track. This would add a parameter (the angular location of the progenitor along the track), and likely introduce a new set of degeneracies. The method will therefore be more computationally expensive, as we would have to explore many more orbits. However, this additional complexity will most clearly impact our inference of the progenitor properties, but to a lesser extent the radial profile of the host.
    
    {\bf Distance to the host galaxy}: We have assumed that we are in the reference frame of the host galaxy, where $x,z$ is the sky plane and $y$ is the line-of-sight direction towards the observer. To apply our method to observations we need to assume a host galaxy distance to transform into galactocentric coordinates. This will affect the scale radius of the system, though we only find weak constraints on the scale radius. The physical properties of each stream orbit will also depend on the distance to the host. If we overestimate the distance to the host, we have in turn over estimated the projected distance from the progenitor to the center of its host in physical units. To match the length of the observed stream, we will need a more massive dark matter halo. However, the width of a stream is roughly proportional to $M_{\rm enc}(r)^{-1/3 }$ (e.g., \citealt{Erkal2016}). Therefore, to match the observed width of a stream at a larger distance from the host a less massive halo would be required. However, width is only a weak function of radius while length is more sensitive to enclosed mass, so we still expect a distance overestimate to bias the inferred mass high. It is possible to include distance uncertainties in our inference, though we defer this to future work.

 {\bf Mock observing the streams}:
    In this paper, we applied an ``observational'' threshold criteria (see \S\ref{sec:methods}), but otherwise assumed perfect data. In observations, we likely only observe the densest parts of streams, which will play into the estimated width of each stream from the control points. This could be addressed by adding uncertainties or scatter to our control points representing each stream (see Fig.~\ref{fig: Histogram_Stream}). We do, however, ensure that streams \emph{longer} than the input data are not disfavored in our sampling.
   
    In halos where two streams are observed, it can be difficult to disentangle whether the two streams are indeed distinct or from the same progenitor (see e.g., \citealt{fielder2025,2026ApJ...997..153B}).
    \citet{miro2024b} found that if we reach a surface brightness limit of 31 mag arcsec$^{-2}$  stellar tidal streams should be detected around 50\% of surveyed hosts, however the occurrence of two streams around one host galaxy might be low.
    The Roman wide field of view combined with its depth will be ideal for such studies. 
    We also note that streams at smaller galactocentric radii, such as the inner stream, could be more difficult to detect due to a lower surface brightness contrast to the stellar halo. While the detection of multiple dwarf galaxy streams around one host galaxy might be rare and difficult to disentangle, our method can also be used on extragalactic globular cluster streams,  predicted to be discovered by Roman \citep{pearson2019, pearson2022a}, or on dwarf streams around dwarf galaxies \citep[e.g.][]{delgado2012,kadofong2018,carlin2019,fielder2025}.

    {\bf Time-dependence}: Throughout this paper, we have assumed that the host potential and progenitor mass are static. From the Milky Way, we know that time dependence induced from accreted satellites \citep[e.g.,][]{erkal2019,Shipp2021,vasiliev2021,2022ApJ...939....2A,dillamore2022,Lilleengen2023,2024ApJ...969...55N,brooks2025} and from the overall mass growth of host halos \citep{buist2015} can affect the orbits and morphologies of stellar streams. 
    However, depending on the location of the stream in its host halo, the recent merger history, and the dynamical age of the streams, some streams may  be less sensitive to time dependence \citep{nibauer2022, brooks2025}. 
     \citet{buist2015} showed that for a smooth halo growth that well reproduced the average mass accretion history of the Milky-Way-sized halos, time evolution led to an angular difference in apocenters at the level of $\approx 10~\deg$, but only for sufficiently long streams which wrapped around their host more than once. 
    In an upcoming work, we test our method on cosmologically evolved halos, where we can determine the bias induced by realistic time-dependence.

{\bf Spherical halo assumption}: We have assumed a spherical potential to represent our host halos.  
$\Lambda$CDM predicts that dark matter halos are often triaxial \citep[][]{frenk1988,dubinski1991,warren1992,jing2002,vera2011,2025ApJ...988..190A}, though more spherical profiles in the inner regions are possible depending on the mass of the baryons \citep{kaza2010}. 
Non-spherical halos will allow for orbital plane precession \citep{Erkal2016, 2024ApJ...969...55N}, which is in turn degenerate with the precession of apocenters in halos with different density slopes.  \cite{nibauer2023} showed that the curvature of stellar streams can be used to infer the flattening of halos, and their (mis)alignment with a disk component. It is possible to use the method of \cite{nibauer2023} to rapidly determine which streams prefer spherical geometries, and target those candidates using the method presented here. Alternatively, it is possible to include flattening in our sampling procedure, though the runtime of our generative method is on the order of hours, while the method of \citet{nibauer2023} works in seconds. In future work, we intend to use the curvature method to inform prior distributions for \texttt{X-Stream}.

{\bf Statistical Assumptions:} Our method employs an average likelihood function rather than the true likelihood. This produces better sampling efficiency and is related to the sampling method of likelihood tempering with a constant temperature. To derive significance regions that enclose 68 and 95\% of the true posterior mass (rather than the tempered density), we calculate contours based on the assumption of normality. We show in Appendix~\ref{app: test_statistic} that this assumption can introduce moderate bias when drawing contours; however, for heavy-tailed distributions, our contours are conservative (i.e., wider than the true posterior). Nevertheless, we have validated that the method produces stream models consistent with the input data within the 68\% confidence regions (e.g., Fig.~\ref{fig:1sigma3sigma}).
    
{\bf Extending to 1000s of halos}: Our nested-sampling technique is limited by its run time ($\approx$ 30 mins for the outer stream, and 1 hour for the inner stream running on 1 GPU). This is not yet scalable to 1000s of galaxies. However, the method can be parallelized across many hosts, and it is also possible to use a more efficient stream generator \citep[e.g., streakline:][]{kuepper2012,kuepper2015}.

\section{Summary \& Conclusion}\label{sec:conclusion}
We present \texttt{X-Stream}, a generative method for constraining gravitational potentials using images of extragalactic stellar streams. The method employs nested sampling with likelihood tempering to efficiently explore the inherently multimodal parameter space involved in inferring halo properties from stream morphology. Simulations are accelerated by generating hundreds of stream models in parallel on a single GPU. Our main conclusions are summarized below:

\begin{itemize}
\item For two distinct streams evolved within a NFW dark matter halo, our method successfully recovers the true orbital parameters using only the projected on-sky morphology. It also accurately infers the progenitor mass, halo mass, and the inner and outer slopes of the dark matter halo's density profile.

\item Extragalactic streams orbiting at different galactocentric radii are sensitive to different regions of the host halo's radial density profile. Consequently, imaging multiple streams enables reconstruction of the full radial profile—from within the scale radius out to the virial radius.

\item We identify new degeneracies between the halo mass and inner and outer density slopes that our method is equipped to account for. We also recover previously known degeneracies for extragalactic streams, including those between radial velocity and line-of-sight position, as well as between progenitor mass, halo mass, and the stream's dynamical age  \citep[see][]{johnston2001,Fardal2013,pearson2022b}.

\item  For a stream evolved in a host galaxy with a cored inner dark matter density profile and a NFW-like outer profile--consistent with predictions from SIDM and fuzzy dark matter--our method accurately recovers the true density slope and disfavors a pure NFW profile at the $2\sigma$ level.
\end{itemize}

Upcoming and ongoing surveys including {\it Euclid} \citep{racca2016}, The Rubin Observatory \citep{Ivezic2008,ivezic2019}, {\it Roman} \citep{spergel2015}, and ARRAKIHS \citep{guzman2022} are expected to discover 1000s of extragalactic stellar streams. The method presented here provides a generative approach to transform stream images into constraints on the underlying mass profile of dark matter halos. Because streams often reside in the outer regions of dark matter halos, the method provides independent constraints on halo shapes and mass limits, and probes a region of the halo that is inaccessible to traditional techniques using stellar and HI kinematics. The method is applicable at the level of individual galaxies, while most previous applications have been limited to galaxy groups or clusters, utilizing X-ray emission \citep[e.g.][]{Reiprich2013} or weak-lensing \citep[e.g.][]{Umetsu2011}. Similar to lensing, it is possible to combine information across many streams or hosts to obtain a higher confidence constraint on halo profiles.
Combined with recent work showing that the shapes and masses of dark matter halos can be inferred from streams \citep{Fardal2013,pearson2022b,nibauer2023,walder2025}, and that gaps from dark matter subhalos can be observed in extragalactic globular cluster tidal tails \citep{aganze2024}, extragalactic stream imaging will offer a powerful test of dark matter across 1000s of galaxies spanning a wide range of masses and redshifts.

\begin{acknowledgements}
This project was developed in part at the Streams24 meeting hosted at Durham University. We thank Adrian Price-Whelan, David Spergel, Nathaniel Starkman, and Oren Slone for insightful discussions.
JN is supported by a National Science Foundation Graduate Research
Fellowship, Grant No. DGE-2039656. Any opinions, findings, and conclusions
or recommendations expressed in this material are those of the author(s) and
do not necessarily reflect the views of the National Science Foundation.
This work was supported by a research grant (VIL53081) from VILLUM FONDEN. 
This work was co-funded by the European Union (ERC, BeyondSTREAMS, 101115754). Views and opinions expressed are, however, those of the author(s) only and do not necessarily reflect those of the European Union or the European Research Council. Neither the European Union nor the granting authority can be held responsible for them.
The Tycho supercomputer hosted at the SCIENCE HPC center at the University of Copenhagen was used for supporting this work. The simulations presented in this article were performed, in part, on computational resources managed and supported by Princeton Research Computing, a consortium of groups including the Princeton Institute for Computational Science and Engineering (PICSciE) and the Office of Information Technology's High Performance Computing Center and Visualization Laboratory at Princeton University.
\end{acknowledgements}

\software{astropy \citep{2013A&A...558A..33A,2018AJ....156..123A}, streamsculptor \citep{Nibauer2025a}, nautilus \citep{nautilus}, JAX  \citep{Jax2018}, Gala \citep{gala}.}

\begin{appendix}

\section{Precession of Apocenters}\label{sec:theory}

\begin{figure*}
\centering\includegraphics[scale=.6]{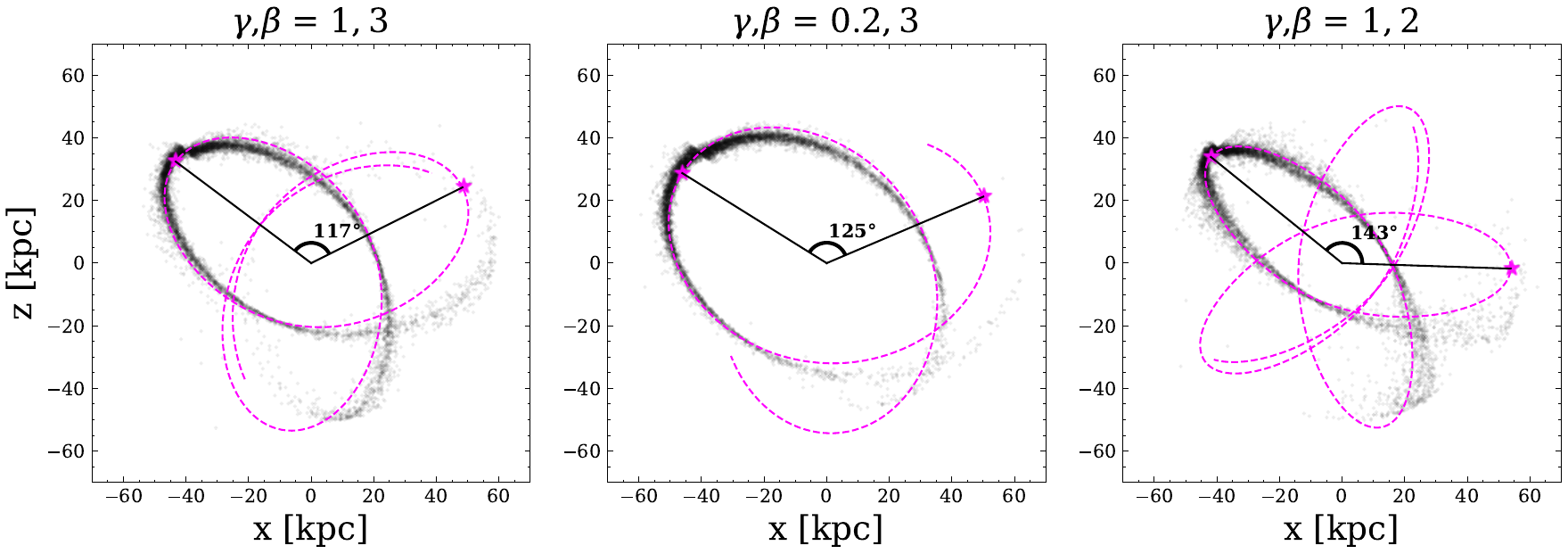}
    \caption{Same as top row of Fig.~\ref{fig: Inner_Density_TwoStream}, but now the inner streams are rotated such that their orbital plane is aligned with the sky plane in ($x,z$). The magenta orbits show 600 Myr of past evolution and 600 Myr of future evolution of the progenitors. The magenta stars show the two most recent apocenters. The progenitors are moving in the positive ($x,z$) direction. We illustrate the apocenter precession angle, $\theta_{\rm apo}$ for all three cases. $\theta_{\rm apo}$ is smallest for higher $\gamma$ and $\beta$, and follows the expected behavior from Fig.~\ref{fig: intuition}. Note that due to longer dynamical times in the cored profile (middle), it takes the progenitor $\approx$600~Myr to traverse from one apocenter to the next. }
    \label{fig:orbitalplane}
\end{figure*} 

We will follow a similar approach to \citet{1998ApJ...495..297J}, and assume near circular orbits allowing us to use the epicycle approximation. The azimuthal period for a circular orbit is
\begin{equation}
T_{\psi} = 2 \pi \sqrt{ r \left( \frac{d\Phi}{dr} \right)^{-1}}.
\end{equation}
The radial period for a near circular orbit is
\begin{equation}
    T_r = \frac{2\pi}{\kappa},
\end{equation}
where $\kappa$ is the epicyclic frequency, defined as
\begin{equation}
    \kappa^2 = \frac{d^2\Phi}{dr^2} + \frac{3}{r} \frac{d\Phi}{dr}.
\end{equation}
The ratio $T_r / T_\psi$ provides a measure of the azimuthal angle swept out by an orbit over a radial period. The angle is
\begin{equation}\label{eq: Delta_Psi}
    \Delta \psi = 2\pi \frac{T_r}{T_\psi}.
\end{equation}
This is the angle between successive pericenters.

The frequency of precession involves the quantities defined above, giving us
    \begin{equation}\label{eq: omega_p}
        \Omega_p = \frac{\vert \Delta \psi - 2\pi \vert}{T_r} = \frac{2\pi}{T_r}\Big\vert \frac{T_r}{T_\psi} - 1\Big\vert.
    \end{equation}
Computing Eq.~\ref{eq: omega_p} as a function of halo inner density slope provides us with a means to make predictions for which orbits are sensitive to different choices of the halo radial profile. 

The time elapsed between two apocenters is itself a radial period. Therefore, the angle between successive apocenters is 
\begin{equation}\label{eq: theta_apo}
    \theta_{\rm apo} = 2\pi \Big\vert \frac{T_r}{T_\psi} - 1\Big\vert.
\end{equation}

In Fig.~\ref{fig:orbitalplane}, we show the three inner streams rotated to their orbital plane and visualize the two most recent apocenters (red stars). $\theta_{\rm apo}$ follows the expected behavior from Fig.~\ref{fig: intuition}, where the stream evolved in a potential with the steepest inner and outer slopes exhibits the smallest value of $\theta_{\rm apo}$.

\section{Validation of Test Statistic}\label{app: test_statistic}
\begin{figure}
\centering\includegraphics[scale=.65]{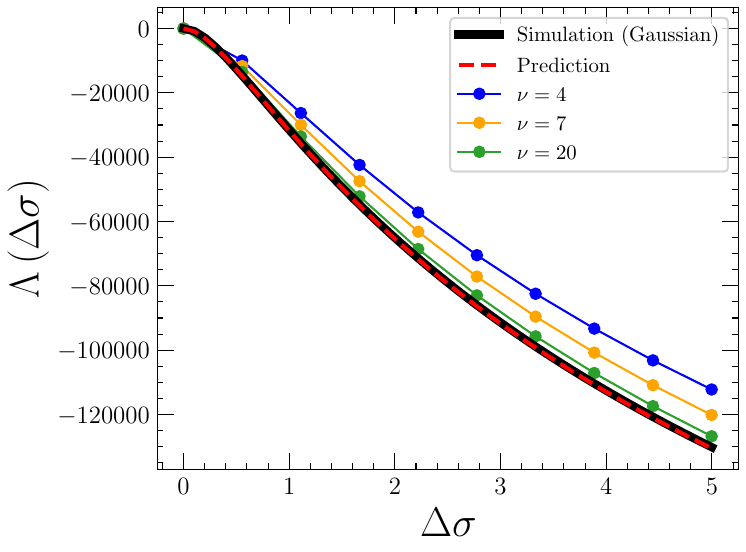}
    \caption{Validation that the derived $\bar{\Lambda}$ from Eq.~\ref{eq: Lambda_bar}  (dashed red curve) matches $\Lambda$ values measured from many simulations (solid black curve) under the assumption of normality. We also compute $\Lambda\left(\Delta \sigma\right)$ for a Student's t-distribution with $\nu$ degrees of freedom. At small $\nu$ (heavier tails) the true $\Lambda$ values are higher than the Gaussian prediction.}
    \label{fig: test_stat}
\end{figure} 

Here we validate the test statistic, $\Lambda$, derived in \S\ref{sec: contours}. Under the assumptions of normality discussed in \S\ref{sec: contours}, we generate samples from two normals and measure the test statistic $\Lambda$ given by Eq.~\ref{eq: lambda_ts}. We perform simulations as a function of the parameter $\sigma = \Delta\sigma + 1$. We than compute analytically Eq.~\ref{eq: Lambda_bar}, and compare the predicted values of $\Lambda$ to the derived values from the simulations in Fig.~\ref{fig: test_stat}. Indeed, there is excellent agreement between the simulated (black) and predicted (dashed red) curves under the assumption of normality. To illustrate departures from a Gaussian distribution, we utilize a Student's t-distribution with $\nu$ degrees of freedom. We consider $\nu \in \{4, 7, 20\}$, noting that as $\nu \xrightarrow{} \infty$, the distribution approaches a standard normal. For each choice of $\nu$, we generate samples and adjust the scale parameter, $s$, such that $x \xrightarrow{} x/s$. For every combination of $\nu$ and $s$, we compute the area under the curve from the center of the distribution up to the 34th percentile. This allows us to derive the $\Delta \sigma$ factor relative to the baseline case where $s = 1$. Finally, we compute the statistic $\Lambda$, with the resulting values illustrated in Fig.~\ref{fig: test_stat} (blue, orange, and green curves for $\nu = 4, 7,$ and $20$, respectively). At $\nu = 4$, the Student's t-distribution is heavy-tailed compared to the normal distribution, resulting in a $\Lambda$ that is larger than the prediction under normality (dashed red curve). As $\nu$ increases, this discrepancy lessens. The discrepancy between the true $\Lambda(\Delta\sigma)$ curves and the normal prediction is maximal at large $\Delta \sigma$. However, in this work, we draw contours using $\Lambda(\Delta \sigma < 2)$. Fig.~\ref{fig: test_stat} illustrates that at this level, our contours can be viewed as conservative, since heavy-tailed distributions produce more positive $\Lambda$ values than a Gaussian distribution at a fixed $\Delta \sigma$.

\section{Outer Stream Constraint}\label{sec:outer}
\begin{figure*}
\centering\includegraphics[scale=.35]{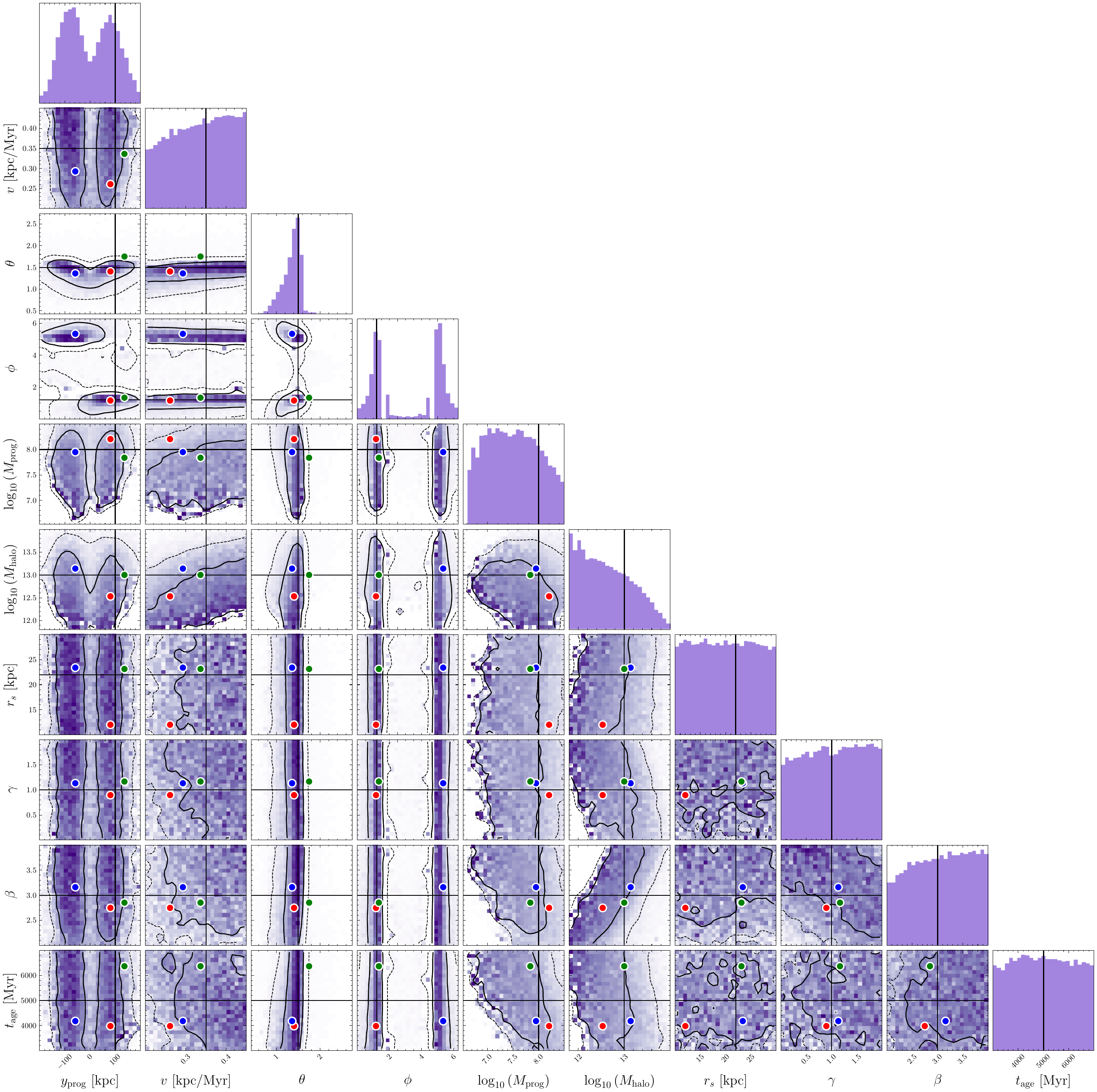}
    \caption{Same as Fig.~\ref{fig:Corner}, but for the outer stream. 
    }
    \label{fig:Corner_outer}
\end{figure*}

In Fig.~\ref{fig:Corner_outer} we present the results of our sampler for the outer stream.  The purple contours show the 68\% and 95\% confidence regions. The true parameters are over plotted as black lines. The colored points represent solutions which fall within the $68\%$ region (blue and red), and a discrepancy solution (green) (see Fig.~\ref{fig:1sigma3sigma} for the corresponding stream models). 
All 10 free parameters for the progenitors and the dark matter halos are recovered within the $68\%$ region, though the stream morphology is not sensitive to the scale radius, $r_s$,  where the posterior distributions are flat, and we can only place a lower limit on the dynamical age of the stream, $t_{\rm age}$. 
Compared to the inner stream (see Fig.~\ref{fig:Corner}), the outer stream places a weaker constraint on the inner density slope, $\gamma$, but a stronger constraint on the outer density slope, $\beta$ (see the discussion of Fig.~\ref{fig: DensityProfile_Constraint} in \S\ref{sec:results}).

\section{Cored Stream Constraint}\label{sec:app_core}
\begin{figure*}
\centering\includegraphics[scale=.45]{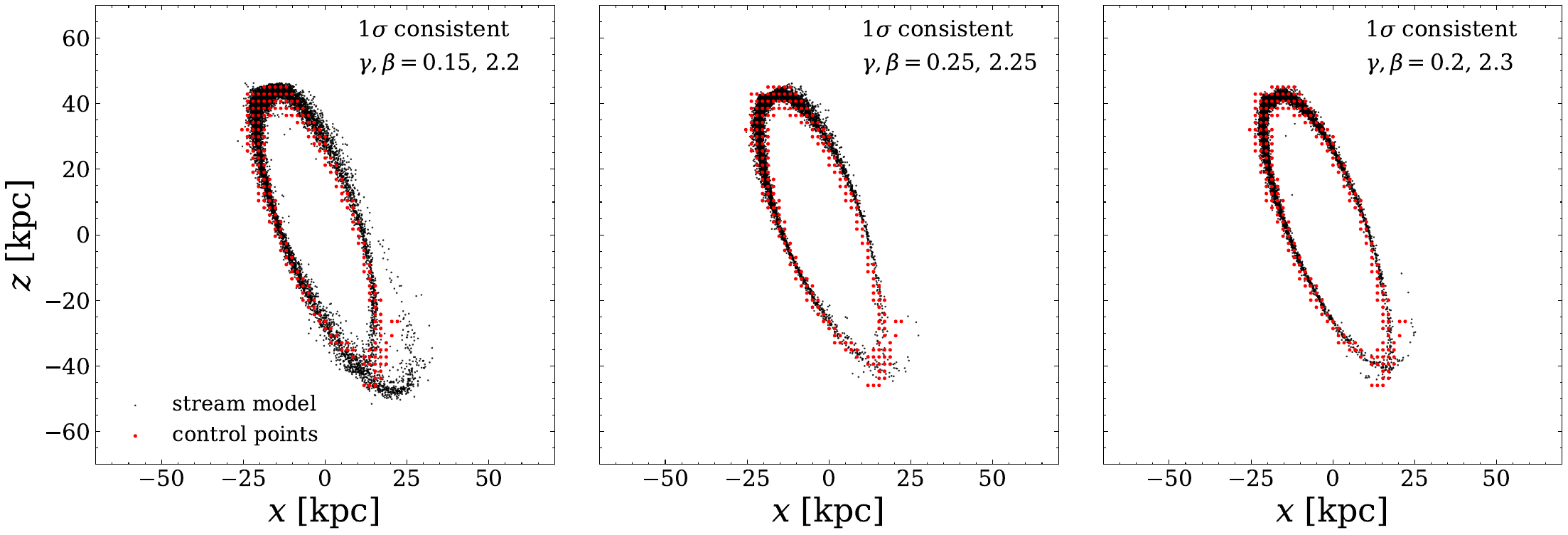}
    \caption{ Three different streams (black points) which are evolved in a halo with low $\gamma$ values (see legends) selected within the 68\% contours in 10 dimensions from Figure \ref{fig: inner_cored}. The red points show the
input data used to represent the inner stream, and all streams produce good fits to the input data. 
    }
    \label{fig:app_cored}
\end{figure*}

In Figure \ref{fig: inner_cored}, the sampler recovers the preference for a cored profile and the true parameters are within 68\% contours. To explore valid fits, we select three different model streams within the 68\% contours from Figure \ref{fig: inner_cored} in the 10 dimensional space, which have $\beta$ = 2.2, 2.25, 2.3 and $0.15<\gamma<0.3$. Note that the other 8 parameters are selected at random within 68\% contours and vary in each case. We visualize these three model streams in Figure \ref{fig:app_cored}. While all streams produce good fits to the input data, there are some differences between each stream. The left stream with the lowest $\gamma$ and $\beta$ is slightly longer and wider than the other streams. This stream also happens to be from a more massive progenitor fit within the  68\% contours compared to the middle and right stream.

\clearpage

\end{appendix}

\bibliography{thebib}
\end{document}